\documentclass[aps,prd,superscriptaddress,preprintnumbers]{revtex4-1}

\usepackage[squaren]{SIunits}
\usepackage{amsmath}
\usepackage{amssymb}
\usepackage{graphicx}
\usepackage{subfig}

\begin{document}

\newcommand{\der}{\text{d}}
\newcommand{\mean}[1]{\left\langle#1\right\rangle}
\newcommand{\fulljust}{\setlength{\rightskip}{0pt}\setlength{\leftskip}{0pt}}

\title{Extragalactic and galactic gamma-rays and neutrinos from annihilating dark matter}
\author{Rouzbeh Allahverdi}
\affiliation{Department of Physics and Astronomy, University of New Mexico, Albuquerque, New Mexico 87131, USA}
\author{Sheldon Campbell}
\author{Bhaskar Dutta}
\affiliation{Department of Physics and Astronomy, Texas A\&M University, College Station, Texas 77843, USA}

\preprint{MIFPA-11-44}

\begin{abstract}
We describe cosmic gamma-ray and neutrino signals of dark matter annihilation, explaining how the complementarity of these signals provides additional information that, if observable, can enlighten the particle nature of dark matter. This is discussed in the context of exploiting the separate galactic and extragalactic components of the signal, using the spherical halo model distribution of dark matter. We motivate the discussion with supersymmetric extensions of the standard model of particle physics. We consider the minimal supersymmetric standard model (MSSM) where both neutrinos and gamma-rays are produced from annihilations. We also consider a gauged $B-L$, baryon number minus lepton number, extension of the MSSM, where annihilation can be purely to heavy right-handed neutrinos. We compare the galactic and extragalactic components of these signals, and conclude that it is not yet clear which may dominate when looking out of the galactic plane. To answer this question, we must have an understanding of the contribution of halo substructure to the annihilation signals. We find that different theories with indistinguishable gamma-ray signals can be distinguished in the neutrino signal. Gamma-ray annihilation signals are difficult to observe from the galactic center, due to abundant astrophysical sources; but annihilation neutrinos from there would not be so hidden, if they can be observed over the atmospheric neutrinos produced by cosmic rays.
\end{abstract}

\maketitle

\section{Introduction}

Understanding the particle nature of dark matter is a major problem of modern physics. While indirect gravitational evidence of its presence is plentiful, unambiguous identification of its particle properties is actively sought. Presumably, this population of particles is a relic that was produced spontaneously in, and is left over from, the big bang. It is common for extensions of the standard model that contain a dark matter candidate to have a $\mathbb{Z}_2$ symmetry that stabilizes the particle from decay, but does not prevent the particle from self-annihilating. Often in viable theories, the annihilations in the early Universe are important to bring the dark matter's relic density down to the observed density today. Once the rate of expansion of the Universe becomes larger than the dark matter annihilation rate, annihilations become rare, and the dark matter relic is said to freeze out at this time. However, if dark matter annihilates, then rare annihilations continue to occur today, predominantly in the densest regions of the Universe. Unambiguous identification of cosmic radiation from these annihilations would not only provide valuable information about the particle nature of dark matter, but also information about the distribution of the matter responsible for the signal.

The information about dark matter from indirect detection of its cosmic annihilation radiation is complementary to that gleaned from other current experiments. Indirect detection experiments constrain the dark matter particle mass, annihilation cross section, and annihilation spectrum. Meanwhile, direct detection experiments attempt to observe dark matter-nucleon interactions in a laboratory detector, and constrain the dark matter mass and its nucleon-scattering cross section. Particle accelerators try to detect dark matter production from particle collisions, where the dark matter would be manifested in missing transverse energy. Again, events of this kind provide information on the dark matter's particle mass, but also on the processes that led to the creation of the dark matter. The results from these experiments provide different constraints for particle physics models, and provide consistency checks for one another.

Today, indirect detection experiments are looking for high-energy cosmic rays, gamma rays, and neutrinos produced from dark matter annihilation. The propagation of produced cosmic rays within the galaxy is difficult to describe precisely, complicating the prediction of the signal's properties. Also, energy losses do not allow extragalactic cosmic rays to reach us. However, neutrinos and gamma-rays can be observed from their sources, both galactic and extragalactic. This allows their observed signals to probe not only the particle physics that produced them, but also the distribution of their sources: the matter in our galaxy or extragalactic large scale structure. A sample of current experiments is the Fermi-LAT \cite{Atwood:2009ez}, currently examining cosmic gamma-rays, and IceCube \cite{Abbasi:2010ak}, which is already monitoring high-energy cosmic neutrinos. In their mandates, they both have commitments to analyze their data for the presence of dark matter annihilation radiation \cite{Baltz:2008wd,*Ackermann:2011bg,*Abdo:2010ex,*Abdo:2010dk,*Ackermann:2010rg,*collaboration:2011wa,*Cuoco:2010jb,*Ajello:2011dq,Abbasi:2009vg,*Abbasi:2011eq}. In this paper, we consider particular models from the minimal supersymmetric standard model (MSSM) where both neutrinos and gamma-rays are produced from annihilations; however, we also discuss different well-motivated models where neutrinos may be the dominant final states. These scenarios would be distinguishable from one another if information were received about dark matter annihilation from observations of both cosmic gamma-rays and neutrinos.

Constraints on dark matter annihilation exist from observations of gamma-rays from dwarf galaxies and neutrinos from the Sun. It is apparent from the lack of an obvious signal from those observations that any existing indirect signal is faint. Successful observation of this radiation, and proper constraints from the observations, will require precise knowledge of other astrophysical radiation sources, and the development of rigorous predictions of the annihilation signal for given realistic models of particle physics and matter distribution. These predictions will be instrumental in understanding which kinds of models are and are not observable, and what constraints will be determined by a given designed experiment. In addition, this research is important for the understanding of precisely what information is available in a given particle model's signal, if observed by a particular experiment. This will not only aid the analyses of current experiments, but will also inform the design of future experiments.

Implementing the dark matter distribution in terms of semi-analytic models allows the basic properties of the densest regions (the cores of halos) to be well-represented. It is reasonable to understand the predictions produced by the simplest models first, learning about the physical scales most important to the signals. Additional features of the distribution, as seen in simulations, can be implemented in the context of these models, and their effects quantified---for example, different models of halo core densities, implementation of halo substructure, implementation of distributions of more complex halo shapes, more general halo statistics, etc. It is usefule to identify the relevant physical scales that determine the main properties of the annihilation signals. This is an important tool to quantify the robustness of the predictions to uncertainties of our knowledge of large scale structure, and tells us about the constraints available to be gained from the experiments. This will also guide our understanding of the annihilation signals calculated directly from simulation data, as in \cite{Zavala:2009zr}.

We share preliminary results of predictions of the intensity spectrum of annihilation gamma rays and neutrinos produced from within and outside the galaxy for a spherical halo model distribution. In Section~\ref{sec:models}, we describe some of the particle physics models that motivate our discussion. We explain the calculation of the observed spectrum of gamma rays from annihilations and show results of the calculations in Sec.~\ref{sec:gamma}. Of particular interest is how different theories can produce the same gamma-ray spectra, even though they have different annihilation modes. We show corresponding results for neutrinos in Sec.~\ref{sec:neutrino} and discuss reasons for their consideration in addition to gamma-rays, such as how they break the degeneracy in spectrum for different models. We provide discussion of the results in Sec.~\ref{sec:discussion}.

\section{Particle Physics Models}
\label{sec:models}

In this paper, we discuss particle models that are supersymmetric extensions of the standard model. In these models, R-parity, or some similar parity property, allows only an even number of supersymmetric partner particles to interact on a fundamental interaction vertex. This stabilizes the lightest supersymmetric particle, which becomes the dark matter candidate.

\subsection{MSSM}

In the MSSM, the particle content is restricted to the standard model, a supersymmetric partner for each standard model degree of freedom, and additional Higgs fields. The supersymmetric charged higgsinos and gauginos mix to produce mass eigenstates called charginos, and the neutral higgsinos and gauginos produce neutralinos. In addition to providing a dark matter mass candidate, this model also stabilizes radiative corrections to the Higgs mass, and causes unification of the standard model forces at an energy called the grand unified theory (GUT) scale.

The most general allowed parameter space for this model has over 100 free parameters. We will restrict our discussion to a small portion of these. In minimal supergravity (mSUGRA) or the constrained MSSM (CMSSM), supersymmetric masses are unified at the GUT scale with scalar masses having value $m_0$, and $m_{1/2}$ being the mass of the gauginos \cite{Chamseddine:1982jx, *Barbieri:1982eh, *Hall:1983iz, *Nath:1983aw, *[{For a review, see }]Nilles:1983ge}. Two Higgs fields each gain vacuum expectation values, the ratio of which is specified with the value $\tan\beta$. We will describe the properties of viable thermally produced dark matter models with vanishing soft supersymmetry-breaking trilinear coupling parameters $A_0$, and positive mass parameter $\mu$ that couples the two Higgs superfields in the superpotential. We will be focusing here on universal masses that are not so large as to result in a dark matter particle massive enough to produce significant top quarks from annihilations.

In this three-dimensional parameter space of $m_0$, $m_{1/2}$, and $\tan\beta$, the parameter space is typically broken up into four main regions: the bulk region, the focus point (also known as hyperbolic) region, the co-annihilation region, and the funnel region. In these regions, the dark matter particle turns out to be the lightest neutralino $\tilde\chi_1^0$.

In the bulk region, both $m_0$ and $m_{1/2}$ are relatively small. The neutralino is nearly pure bino (the gaugino which is the supersymmetric partner of the weak hypercharge gauge boson), and annihilates predominantly to bottom anti-bottom quark pairs $b\overline b$, secondarily to tau anti-tau lepton pairs $\tau^+\tau^-$ (moreso at larger $\tan\beta$). These processes in the bulk region give the correct annihilation cross section to account for the relic density, if it were thermally produced.

Generically, larger values of $m_0$ and $m_{1/2}$ result in theories with larger mass dark matter that have smaller annihilation cross sections, and therefore would result in more thermally produced dark matter in the Universe than is observed today. However, when considered carefully, we see other parameter space does result in the correct relic density, due to different mechanisms \cite{Griest:1990kh}, according to the parameter space of interest.

The focus point region \cite{Chan:1997bi, *Feng:1999mn, *Feng:1999zg, *[{see also }]Baer:1995nq, *Baer:1995va, *Baer:1998sz} has a branch where $m_{1/2}$ remains small and $m_0$ is allowed to increase. As $m_0$ does so, the lightest neutralino gains a larger Higgsino component, which opens up additional annihilation channels. Here, annihilation dominantly produces $W^+W^-$ bosons, with a small branching fraction also producing $ZZ$ boson pairs for small to moderate $\tan\beta$. For large $\tan\beta$, the Higgsino component of the lightest neutralino is again small in this region, but Bino annihilation is enhanced by an increased coupling to the pseudoscalar Higgs $A$ and annihilation is again dominated by $b\overline b$ and $\tau^+\tau^-$. Some parts of this parameter space have restrictions from the results of the CMS \cite{Chatrchyan:2011qs,*Chatrchyan:2011zy} and ATLAS \cite{Collaboration:2011iu,*Aad:2011qa} experiments at the Large Hadron Collider.

There is a threshold where $m_0$ becomes too small and one of the supersymmetric partners of the tau (stau $\tilde\tau$) becomes the lightest supersymmetric particle, which is electrically charged and therefore cosmologically disallowed. This threshold increases with $m_{1/2}$. Near this boundary, the $\tilde\tau$ mass is only slightly larger than the $\tilde\chi_1^0$ mass, enhancing the co-annihilation interaction cross section between these particles. The $\tilde\tau$'s present in the early Universe co-annihilate with the $\tilde\chi_1^0$'s, and reduces the neutralino density to the correct value. This parameter space is the stau-neutralino co-annihilation region \cite{Ellis:1998kh, *Ellis:1999mm, *Gomez:1999dk, *Gomez:2000sj, *Lahanas:1999uy, *Arnowitt:2001yh}. When $A_0>0$, there is parameter space at low $m_{1/2}$ where a supersymmetric partner of the top quark (stop $\tilde t$) becomes lighter than $\tilde\chi_1^0$. The stop-neutralino co-annihilation region \cite{Ellis:2001nx} is near this boundary. In these parameter spaces, $\tilde\chi_1^0$ is again nearly pure bino and mostly $b\overline b$ and some $\tau^+\tau^-$ are produced from annihilations. Because there are no $\tilde\tau$ or $\tilde t$ particles present today, they no longer contribute to annihilations and the effective annihilation cross section of the neutralinos is reduced from its value at freezeout. Additionally, at low $\tan\beta$, annihilation is dominated by t-channel sfermion exchange, which is helicity-suppressed \cite{Drees:1992am,*Bell:2010ei}. The presence of a strong p-wave annihilation component brings the annihilation cross section up to its needed value at freezeout, but slow relative motions of the particles today do not allow the p-wave to contribute. In these cases, the annihilation cross-sections are quite small, which make the rates of annihilations low and the intensity of annihilation radiation much more difficult to detect. The situation becomes better at large $\tan\beta$ where annihilation via $A$ is a stronger component, lifting much of the helicity suppression.

The final parameter space, the heavy Higgs or $A$ annihilation funnel regions, occurs where the mass of one of the Higgs bosons is near half the $\tilde\chi_1^0$ mass, resulting in a Breit-Wigner resonance enhancement of the annihilations at freezeout interaction energies \cite{Baer:1997ai, *Baer:2000jj, *Ellis:2001msa, *Roszkowski:2001sb, *Djouadi:2001yk, *Lahanas:2001yr}. Since the resonance does not enhance the cross section today, the annihilation cross sections are again lower in the present epoch for dark matter models of this parameter space.

\subsection{Gauged $U(1)_{B-L}$ Model}
\label{sec:B-L}

Another paradigm we wish to discuss, which is interesting in the context of neutrino radiation production, is the $U(1)_{B-L}$ extension of the MSSM \cite{Mohapatra:1980qe, *[{[Erratum-ibid. }][{]}]Mohapatra:1980qf}. Here, baryon number $B$ minus lepton number $L$ is a gauged charge with associated gauge boson $Z'$ that couples to baryons and leptons, according to their $B-L$ charges with gauge coupling $g'$. This extension requires the presence of right-handed neutrinos $N^c$ for anomaly cancellation, providing a natural framework to explain neutrino masses and oscillations. In order for this new internal symmetry to be spontaneously broken, we must introduce two new Higgs superfields $\mathbf{H}_1'$ and $\mathbf{H}_2'$, standard model neutral and oppositely charged under $B-L$ for anomaly cancellation. They are coupled by a new mass parameter $\mu'$ in a new term added to the MSSM superpotential. The physical neutrinos $\nu$ are light, but $N^c$ heavy, by the type I see-saw mechanism \cite{Minkowski:1977sc, *Yanagida:1979as, *Gell-Mann:1979, *Glashow:1979, *Mohapatra:1979ia}. This requires a Majorana mass for the $N^c$, which does not obey the $B-L$ symmetry; however, the $N^c$ can have a Yukawa coupling to another a Higgs field with lepton number $-2$, which we identify with $H_2'$. This Higgs will gain a vacuum expectation value around $\unit{1}{\tera\electronvolt}$, producing the $N^c$ Majorana mass and generating the appropriate neutrino spectrum. Thus, by defining supersymmetric partners for each of the introduced new fields and putting them in chiral supermultiplets, the minimal $U(1)_{B-L}$ extension to the MSSM has superpotential \cite{Allahverdi:2009ae, *Allahverdi:2008jm, *Allahverdi:2009se}
\begin{equation}
  W=W_{\text{MSSM}}+y_D\mathbf{N}^c\mathbf{H}_u\mathbf{L}+f\mathbf{H}_2'\mathbf{N}^c\mathbf{N}^c+\mu'\mathbf{H}_1'\mathbf{H}_2'
\end{equation}
where $\mathbf{H}_u$ is the Higgs superfield of the MSSM that gives mass to the up-type quarks, and $\mathbf{L}$ is the superfield containing the left-handed leptons. Note that flavor and the weak isospin $SU(2)_L$ indices have been suppressed.

There exists parameter space in this framework where the LSP is a supersymmetric partner of $N^c$, the right sneutrino $\tilde N$. If the $N^c$ mass is less than the $\tilde N$ mass, then annihilations could produce a large number of $N^c$, which would then decay according to the particular model considered. In any case, one would expect many direct neutrinos to be produced, while photons would only be produced secondarily.

The particular model we will consider will be a parameter space where the $\tilde N$ has a mass of $\unit{150}{\giga\electronvolt}$. In this case, the dominant annihilation channels are the s-wave processes $\tilde N \tilde N \longrightarrow N^cN^c$ and $\tilde{N}^{\mbox{*}}\tilde{N}^{\mbox{*}}$ $\longrightarrow N^{c\mbox{*}}N^{c\mbox{*}}$ via t-channel exchange of $B-L$ neutralinos through its coupling with the gaugino $\tilde Z'$. The $N^c$, taken to have mass $\unit{135}{\giga\electronvolt}$, then decay exclusively to $\nu$ and standard model Higgs $h$, which we took to have mass $m_h=\unit{120}{\giga\electronvolt}$. At this mass, the Higgs decays mostly to $W W^*$ bosons, and to $b\overline b$ quarks, each of which produce secondary photons and neutrinos. We will also discuss cases where $\tilde N$ is heavier.

\section{Gamma-ray Results}
\label{sec:gamma}

The production of gamma-rays due to extragalactic dark matter annihilation is estimated using the halo model distribution of dark matter used in \cite{Campbell:2010xc,*Campbell:2011kf}, where each halo is specified by its mass $M$ and observed redshift $z$. The Sheth-Tormen halo mass function $\frac{\der n}{\der M}$ \cite{Sheth:1999su, *Sheth:2001dp} is a good approximation of the halo distributions seen in simulations. We take typical halos to be spherical, with a Navarro-Frenk-White (NFW) density profile \cite{Navarro:1996gj} 
\[
  \rho_h(r|M,z)=\frac{\rho_s(M,z)}{[r/r_s(M,z)][1+r/r_s(M,z)]^2}
\]
extending out to virial radius
\[
  R_\text{vir}(M,z)=\left(\frac{3M}{4\pi \Delta_{\text{vir}}\mean{\rho}\!(z)}\right)^{1/3}
\]
with $\Delta_{\text{vir}}=180$ and $\mean{\rho}\!(z)$ being the background matter density at redshift $z$. The distribution of scale radii $r_s$ is described by the distribution of halo concentrations $c\equiv\frac{R_\text{vir}}{r_s}$, which is approximately described by the physically motivated model of \cite{Bullock:1999he} over the mass scales probed by simulations. We set the minimum halo mass scale at $10^{-6} M_{\odot}$. Cosmological parameters used were from the WMAP7 data \cite{Komatsu:2010fb}, and we use the linear power spectrum of \cite{Eisenstein:1997jh} to describe large-scale fluctuations in the matter distribution. These inputs describe the simulations well enough to give us a good approximation of the annihilation signal. However, as we will establish, the signal (produced from dark matter distributed according to our halo model) is sensitive to low mass halo properties and halo core properties, which are beyond the reach of current simulations, and the knowledge of which will require a good understanding of interactions with baryonic matter. We consider all photons emitted since the epoch of reionization, which we estimate to have occurred at redshift $z_\text{max}=10$.

Given this description of the densest regions of large scale structure, the mean extragalactic intensity of gamma-rays from the annihilation of dark matter particles, each of mass $m$, at constant s-wave relative-velocity-weighted annihilation cross section $\sigma v$ with annihilation spectrum $\frac{\der N_\gamma}{\der E_\gamma}$ is
\begin{equation}
  I_{\gamma,\text{EG}}(E_\gamma)=\sigma v\int\frac{\der z}{H(z)}W((1+z)E_\gamma,z)\mean{\rho^2}\!(z).
\end{equation}
The intensity window function is
\begin{equation}
  W(E_\gamma,z)=\frac{1}{8\pi m^2}\frac{1}{(1+z)^3}\frac{\der N_\gamma}{\der E_\gamma}(E_\gamma)\,e^{-\tau\!(E_\gamma,z)}
\end{equation}
where $\tau(E_\gamma,z)$ is the cosmic opacity to gamma-rays \cite{Stecker:2005qs, *Stecker:2006eh}, and the mean square matter density is determined from
\begin{equation}
  \mean{\rho^2}\!(z)=\int\der M\frac{\der n}{\der M}(M,z)\int\der^3\mathbf{r}\rho_h^2(r|M,z).
\end{equation}

Note that it is possible that $\sigma v$ depends on the relative velocity of the annihilating particles. The most prominent example of this, which appears in particle models, is the presence of p-wave annihilation \cite{Barger:1987xg}. However, this annihilation component must be very large to affect the observed annihilation signal, and thermal freezeout models for which this is true have very small cross sections and are difficult to see \cite{Campbell:2010xc}. Since, in this paper, we are focusing on models where dark matter is thermally produced, it is reasonable to restrict our initial investigations to pure s-wave annihilation. However, if a special case scenario, such as resonant annihilation \cite{Feldman:2008xs, *Ibe:2008ye, *Backovic:2009rw, Hisano:2004ds, *MarchRussell:2008yu, *ArkaniHamed:2008qn, *Lattanzi:2008qa, *MarchRussell:2008tu, *Iengo:2009ni, *Iengo:2009xf, *Cassel:2009wt} were present, then the calculation needs to be modified, as described in \cite{Campbell:2010xc}.

If we take our own Milky Way Galaxy dark matter halo to be a typical halo of our large scale structure model at mass $M_\text{G}=2\times10^{12} M_\odot$, then it has scale radius $r_{s,\text{G}}=\unit{38.0}{\kilo pc}$, virial radius $R_\text{vir,G}=\unit{412}{\kilo pc}$, and concentration $c_{\text{G}}=10.8$. The important parameter here for our calculation is the scale radius, since the contribution to the annihilation signal due to dark matter outside this radius is very small; therefore, the virial radius definition (and hence the value of concentration) does not significantly affect the prediction of the galactic annihilation signal. We estimate the solar system's position in the halo as being $R_\odot=\unit{8.0}{\kilo pc}$ from the galactic center.

With this description, the intensity of gamma-rays, due to dark matter annihilation in the galactic halo in the direction of angle $\psi$ from the galactic center (assumed coincident with the halo center), is typically written as
\begin{equation}
  \label{eq:galinten}
  I_{\gamma,\text{G}}(E_\gamma,\psi)=\frac{\sigma v}{8\pi m^2}\frac{\der N_\gamma}{\der E_\gamma}(E_\gamma) \hat{J}(\psi)
\end{equation}
where the $J$-factor is the line of sight integration of the square dark matter density from the solar system out the halo \footnote{The $J$-factor is usually scaled to be in units of $R_\odot\rho_\odot$, where $\rho_\odot$ is the estimated local density. This is less convenient for comparison with the extragalactic signal, and therefore, we keep the $J$-factor with arbitrary units, and use the symbol $\hat{J}$ to distinguish it from the more standard definition, to prevent confusion.}
\begin{equation}
  \hat{J}(\psi)\equiv\int_0^{r_\text{max}(\psi)}\der r \left[ \rho_{h}\!\!\left(\sqrt{r^2-2rR_\odot\cos\psi+R_\odot^2}\Bigg|M_\text{G},0\right)\right]^2
\end{equation}
with
\begin{equation}
  \label{eq:galrmax}
  r_\text{max}(\psi)=R_\odot\cos\psi+\sqrt{R_\text{vir,G}^2-R_\odot^2\sin^2\psi}.
\end{equation}

\begin{figure*}
  \subfloat{\includegraphics[width=0.33\textwidth]
    {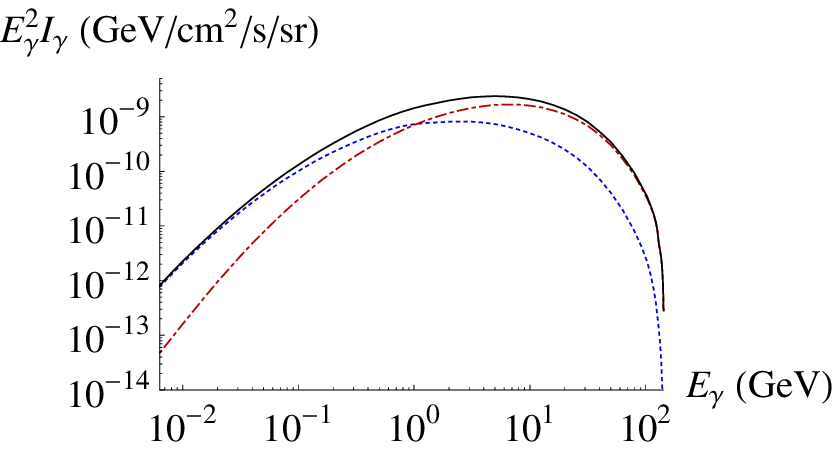}}
  \subfloat{\includegraphics[width=0.33\textwidth]
    {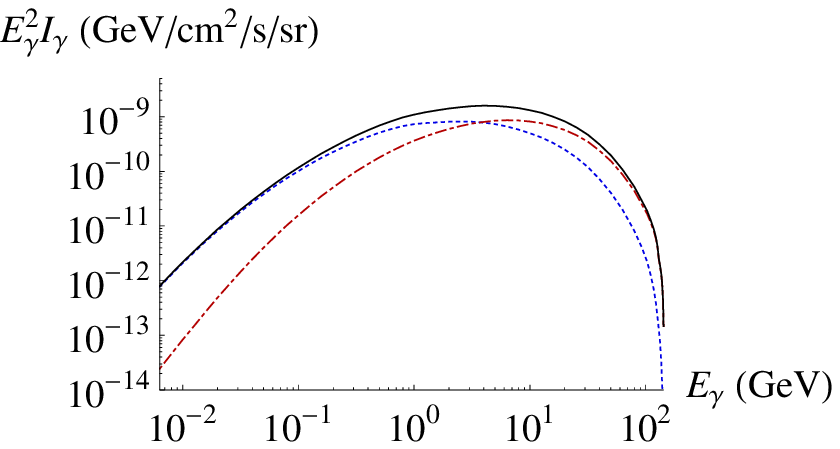}}
  \subfloat{\includegraphics[width=0.33\textwidth]
    {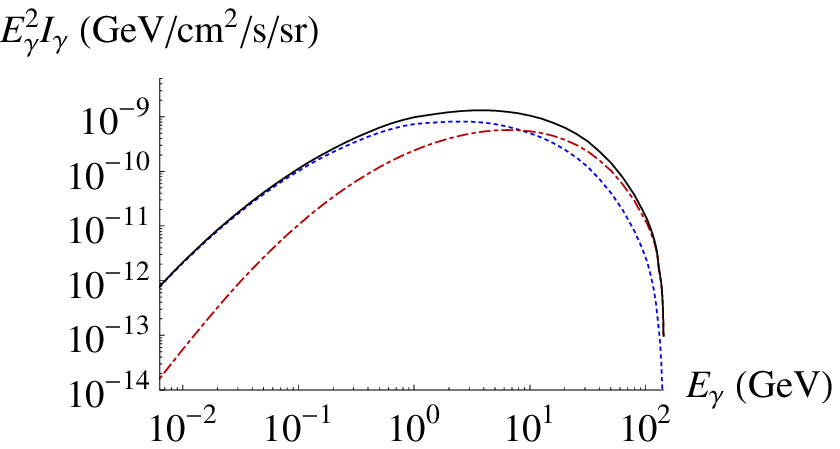}}
  \caption{\label{fig:focuspt_intensities}\fulljust The gamma ray signal from annihilating dark matter in the directions $\psi=30^\circ$, $90^\circ$, and $150^\circ$ from the galactic center, respectively. The dark matter shown here is a $\unit{150}{\giga\electronvolt}$ neutralino in the focus point region of mSUGRA with $\tan\beta=10$. The dotted (blue online) line is the extragalactic component, the dot-dashed (red online) line is the galactic component. The solid line is the net signal. \hfill\mbox{}}
\end{figure*}

In Fig.~\ref{fig:focuspt_intensities}, we see the contributions of the galactic and extragalactic components of annihilation to the gamma-ray intensity for different lines of sight in the halo. The particle physics model used in this example is the focus point region in mSUGRA parameter space where the dark matter particle is the lightest neutralino having mass  $m=\unit{150}{\giga\electronvolt}$ at $\tan\beta=10$, $A_0=0$, and $\text{sign}\ \mu>0$. Here, annihilation is predominantly to $W^+W^-$ boson pairs. The particle annihilation cross section and annihilation spectrum were calculated using the computer program DarkSUSY~5.0.5 \cite{Gondolo:2004sc}. The cross section for this model is $\sigma v=\unit{1.9\times10^{-26}}{\centi\meter\cubed\per\second}$.

\begin{figure*}
  \subfloat{\includegraphics[width=0.4\textwidth]
    {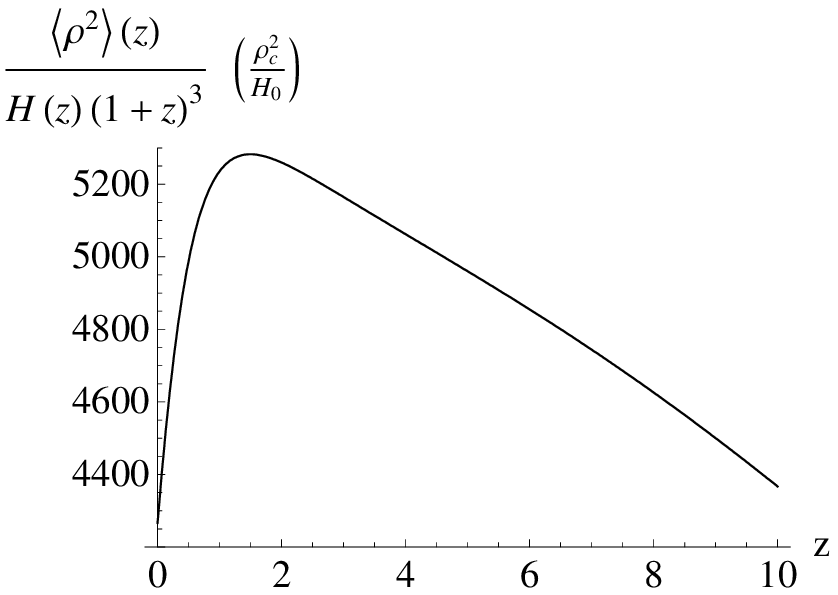}}\quad
  \subfloat{\includegraphics[width=0.4\textwidth]
    {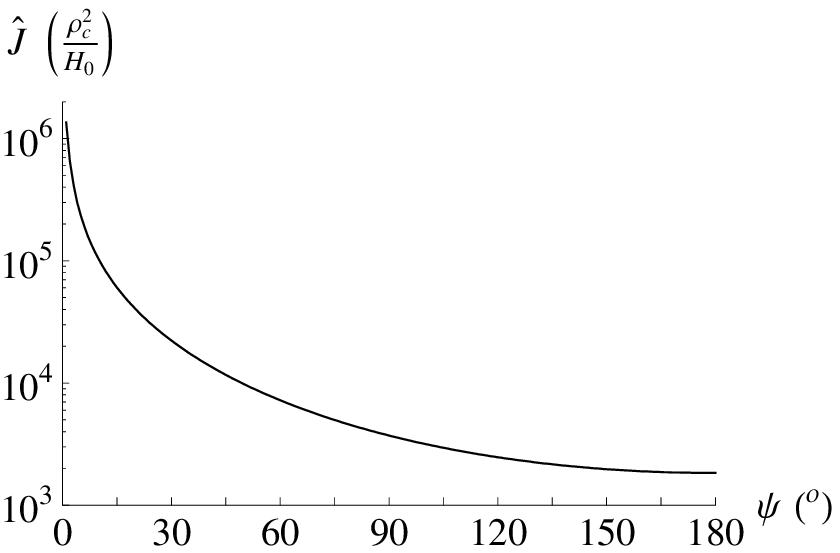}}
  \caption{\label{fig:egtogmags}\fulljust Left: The magnitude of extragalactic intensity is approximately proportional to the area under this curve, around $48 000 \rho_c^2/H_0$. Right: The corresponding contribution to the galactic intensity. \hfill\mbox{}}
\end{figure*}

\begin{figure*}
  \includegraphics[width=0.4\textwidth]{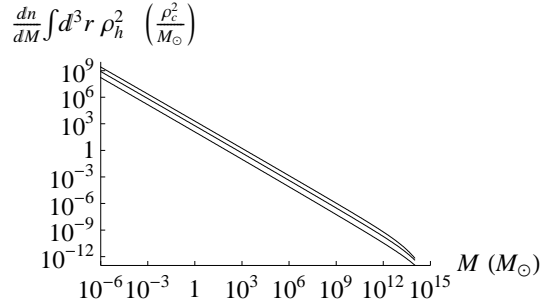}
  \caption{\label{fig:massinten}\fulljust The mass integrand of the mean square density at redshift $z=0$, $0.5$, and $1$ from bottom to top, respectively. \hfill\mbox{}}
\end{figure*}

In our dark matter density distribution models, the galactic component is dominant at the peak of the signal when looking toward the galactic center, but the contributions of the components are comparable when looking out of the galactic plane or away from the galactic center. It is conceivable that with slightly different choices of distribution parameters, the relative importance of each may be altered considerably. The relative strength of the galactic to extragalactic intensity at a given photon energy is
\[
  \frac{I_{\gamma,\text{EG}}(E_\gamma)}{I_{\gamma,\text{G}}(E_\gamma,\psi)}=\int\der z\left[\frac{\mean{\rho^2}\!(z)}{H(z)(1+z)^3\hat{J}(\psi)}\right]\left[\frac{\frac{\der N_\gamma}{\der E_\gamma}((1+z)E_\gamma)}{\frac{\der N_\gamma}{\der E_\gamma}(E_\gamma)}\right] e^{-\tau((1+z)E_\gamma,z)}
\]
The important parameters then appear in the first factor of the integrand. Fig.~\ref{fig:egtogmags} plots the extragalactic and galactic contributions to this factor in units of $\rho_c^2/H_0$, where $\rho_c$ is the cosmological critical density to collapse and $H_0$ is the Hubble constant. The extragalactic part is relatively flat in scale, with an area under the curve of around $48 000 \rho_c^2/H_0$. The convolution with the annihilation spectrum and opacity could modify the importance of this factor, depending on the details of those functions. One may wonder what mass scale of halos most contributes to the mean square density $\mean{\rho^2}$. In Fig.~\ref{fig:massinten}, we can see that the mass integrand goes very nearly like $M^{-1}$ all the way to the maximum mass scale, suggesting that all mass scales contribute nearly equally to the intensity. If the mass dependence of the Sheth-Tormen mass function correctly describes the halo distribution down to low scales, and those low-mass halos have density profiles well-described by NFW, with concentrations described by the model specified, then all mass scales are important contributors to annihilations.

However, let us consider for the moment the effect of the neglected substructure. At Milky Way size halos, it is expected that substructure will increase the annihilation rate by a factor on the order of 100, depending on the minimum halo mass scale \cite{Afshordi:2009hn}. By definition, the smallest halos will not have any subhalos, and larger halos will have more and more substructure. Thus, one would expect the largest halos to contribute the most to intensity purely on the basis of their substructure.

For the galactic contribution, if our galactic halo is well-described by an NFW (or similar) profile, then the value of scale radius $r_s$ has a significant effect on how concentrated the dark matter is to the galactic core. Based on observations of stellar velocities, it is generally estimated that our galactic halo has a somewhat smaller scale radius than the typical radius we used \cite{Klypin:2001xu}. This would result in an increase in the predicted galactic intensity.

The scaling of the density at the core is also important. On the right plot of Fig.~\ref{fig:egtogmags}, we see how the intensity formally diverges as the line of sight approaches the galactic center for the NFW profile. Observing a signal from toward the galactic center would help to better understand how the density is distributed there in our halo, and would allow us to test various ideas about the effects that the central black hole and baryonic cooling have on the profile. 

It is expected that the substructure observed in the simulations would increase the galactic signal by a factor of a few, not as significantly as the extragalactic intensity \cite{Afshordi:2009hn}. Therefore, it is not unreasonable to suppose that the extragalactic annihilation could dominate over the galactic signal for most lines-of-sight that are not too close to the galactic center.

In summary, our estimation of the most crucial elements in these calculations, which have the greatest effects on the result, is:
\begin{itemize}
  \item the halo scale radius, the galactic value of which has an important effect on the galactic signal component, and the halo distribution of which affects the extragalactic signal; and
  \item the inclusion of subhalos, not yet taken into account, will also increase the predicted signal, and will depend on the scale of minimum halo mass.
\end{itemize}
Thus, one can conclude from this discussion that the galactic and extragalactic annihilation signals in Fig.~\ref{fig:focuspt_intensities} are of comparable intensity, due to our value of $r_{s,G}$, and the lack of substructure effects.

\begin{figure*}
  \subfloat{\includegraphics[width=0.4\textwidth]
    {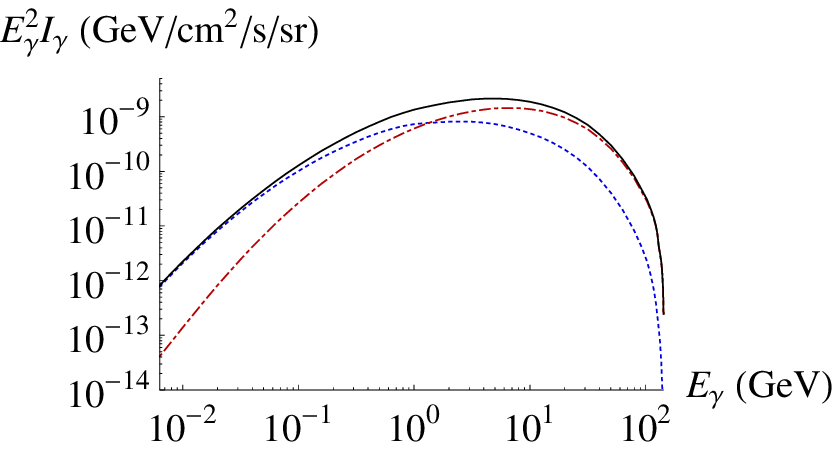}}
  \subfloat{\includegraphics[width=0.4\textwidth]
    {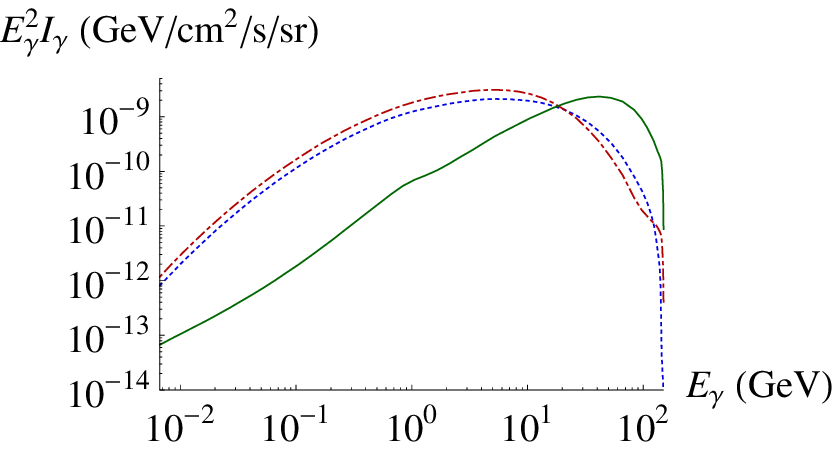}}\quad
  \subfloat{\includegraphics[width=0.1\textwidth]
    {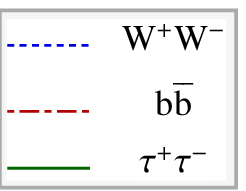}}
  \caption{\label{fig:gamanticore}\fulljust Left: The mean intensity for the focus point model, averaged over all directions $\psi>18^\circ$ away from the galactic center. The plot format is the same as for Fig.~\ref{fig:focuspt_intensities}. Right: The same calculation for a 150 GeV dark matter particle that annihilates purely to $W^+W^-$, $b\overline{b}$, or $\tau^+\tau^-$. \hfill\mbox{}}
\end{figure*}

Unless the annihilation at the galactic core is very bright, it will be difficult to observe those dark matter annihilation gamma-rays originating from there because there are so many other bright sources of astrophysical gamma-rays in that region that have theoretical uncertainties associated with them. A less contaminated signal, for example, would be the consideration of the mean annihilation signal away from the core. The galactic and extragalactic components for this are shown in Fig.~\ref{fig:gamanticore} for the same focus point model. For comparison, we also show total signals for dark matter, of the same mass, that annihilates to $W^+W^-$, $b\overline{b}$, or $\tau^+\tau^-$, at the same annihilation cross section as the focus point model. The annihilation spectra for these models were simulated with the event generator Pythia~6.135 \cite{Sjostrand:2007gs}. The sources of photons in these models are from decaying pions or radiating charged fermions. The W and b spectra are more dispersed to lower energies because they are more likely to decay to hadronic showers where each photon-emitting product is at lower energy. At 150 GeV dark matter annihilation, the photons from annihilation to $W^+W^-$ are indistinguishable from annihilation to $b\overline b$. These pure branching ratio intensities can be used to construct the intensity profile for any theory that annihilates to these states, with known branching ratios. For larger dark matter masses, the $W$ and $b$ signals become more distinguishable from one another.

\section{Neutrino Results}
\label{sec:neutrino}

Because the models discussed in the previous section also contribute a neutrino annihilation spectrum $\frac{\der N_\nu}{\der E_\nu}$, it is interesting to consider this component of the signal as well. Because the neutrino is electrically neutral and weakly interacting, it also propagates relatively freely through the cosmos, and the annihilation signal will have both galactic and extragalactic contributions. This calculation is completely analogous to that for the gamma-ray signal. We neglect any cosmic opacity for the neutrinos in the sample calculations that follow.

\begin{figure*}
  \subfloat{\includegraphics[width=0.33\textwidth]
    {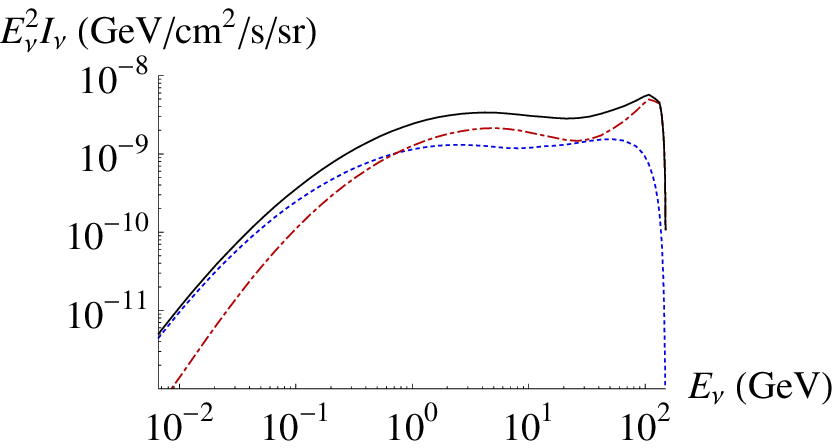}}
  \subfloat{\includegraphics[width=0.33\textwidth]
    {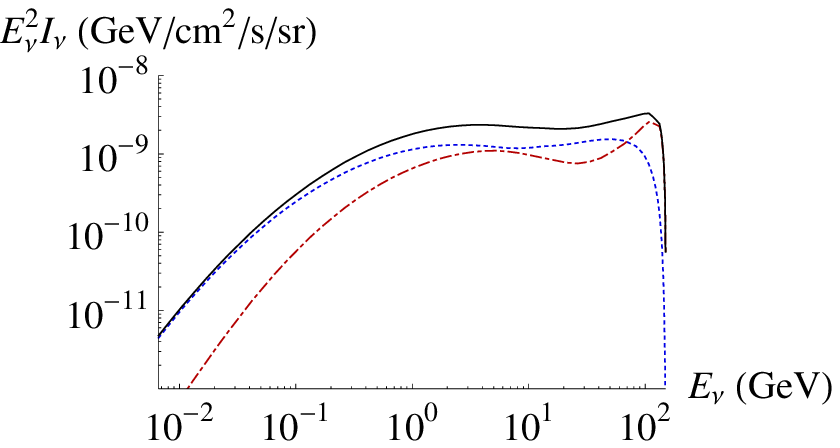}}
  \subfloat{\includegraphics[width=0.33\textwidth]
    {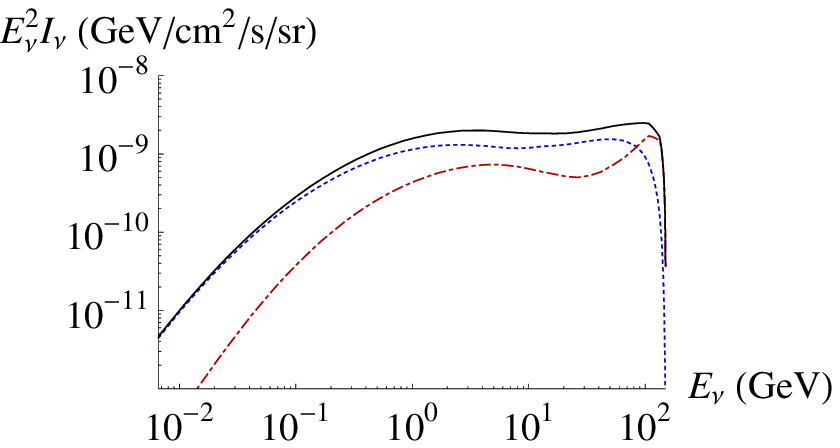}}
  \caption{\label{fig:focuspt_inten_neu}\fulljust The neutrino signal from annihilating dark matter in the directions $\psi=30^\circ$, $90^\circ$, and $150^\circ$ from the galactic center, for the same particle model and plot format as in Fig.~\ref{fig:focuspt_intensities}. \hfill\mbox{}}
\end{figure*}

In Fig.~\ref{fig:focuspt_inten_neu}, we show the galactic, extragalactic, and net intensity of cosmic neutrinos from annihilations of the same 150 GeV focus point neutralino dark matter considered in the previous section. In the galactic signal, we clearly see the peaks from primarily and secondarily produced neutrinos from the W decays. However, those features are washed out in the redshift-modulated extragalactic signal. We note how both components contribute significantly to the total signal in all of the shown lines of sight. Again, reasonable adjustments of dark matter distribution parameters and consideration of halo substructures could significantly alter this balance in either direction.

Although the neutrino signal still suffers from uncertainties in the galactic core density profile, it does not suffer from the same astrophysical contamination as do gamma-rays. Therefore, there is no reason to exclude the galactic center in these experiments. In fact, if a neutrino detector with high angular resolution can be developed, it is a good strategy to focus on the galactic center.

\begin{figure*}
  \subfloat{\includegraphics[width=0.33\textwidth]
    {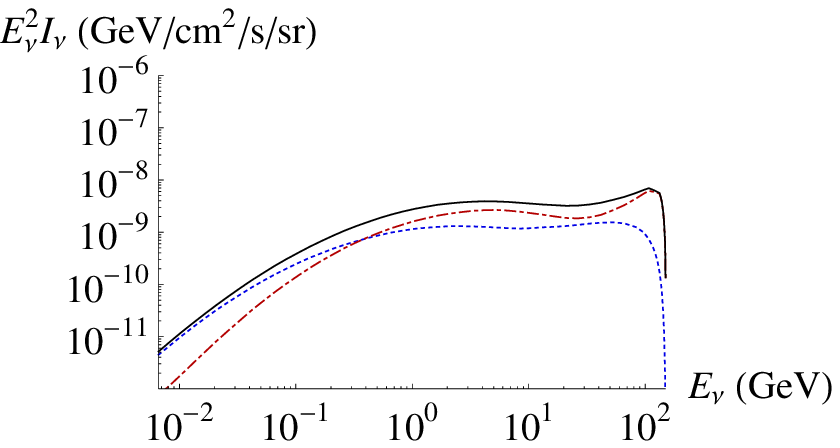}}
  \subfloat{\includegraphics[width=0.33\textwidth]
    {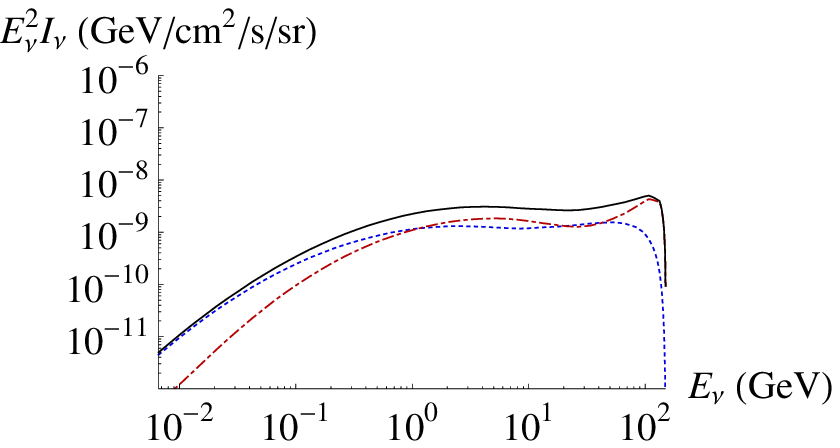}}
  \subfloat{\includegraphics[width=0.33\textwidth]
    {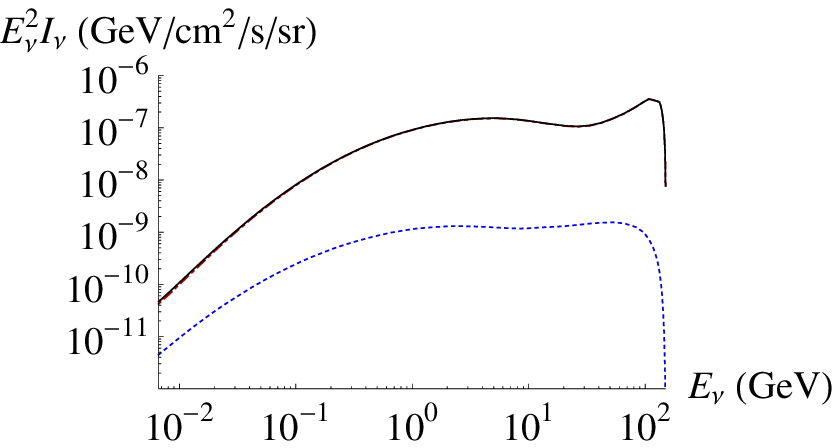}}
  \caption{\label{fig:focuspt_meaninten_neu}\fulljust The mean neutrino intensity for the focus point model. Left: The all-sky intensity, $0^\circ<\psi<180^\circ$. Middle: The anticore intensity, $\psi>18^\circ$. Right: The core intensity, $\psi<5^\circ$. The plot format is the same as in Fig.~\ref{fig:focuspt_intensities}. \hfill\mbox{}}
\end{figure*}

Fig.~\ref{fig:focuspt_meaninten_neu} shows the neutrino signal for the focus point model averaged over the whole sky, directions away from the core, and directions focused on the core, respectively. We see how the galactic signal is seen to dominate the signal at the galactic core if we assume the NFW profile holds to the center, and we neglect extragalactic substructures. The same dominance of the galactic core occurs with annihilation gamma-rays, but it is very difficult to see those photons from the noisy center of the galaxy. Further work, with more realistic distributions, should better elucidate the situation at the galactic core, and provide an understanding of the information about the dark matter distribution uncertainties that may be available in an observed neutrino signal.

It is common in the literature to express neutrino signals as binned detector event rates per detector mass. If $\mean{I_\nu}_\Omega$ is the mean annihilation intensity in a solid angle $\Omega$ of observation, the event rate for a neutrino $\nu_f$ of flavor $f=e$, $\mu$, or $\tau$ in an energy bin $E_i<E_\nu<E_{i+1}$ is
\[
  R_{\nu_f\!,\,i}=\frac{N_A\Omega}{n_m}\int_{E_i}^{E_{i+1}}\der E_\nu\,\sigma_{\nu_f\!N}\!(E_\nu)\mean{I_\nu}_\Omega\!\!(E_\nu)
\]
where $N_A$ is Avogadro's number, $n_m$ is the molar mass of the detector material, and $\sigma_{\nu_fN}$ is the neutrino-nucleon charged current scattering cross section \cite{Gandhi:1998ri, *Jeong:2010nt}. Note that $N_A/n_m$ is simply the nucleon number per detector mass. To ease conversion of the results for different detector materials, we show the results for $n_m=\unit{1}{\gram\per\mole}$.

\begin{figure*}
  \includegraphics[width=0.4\textwidth]{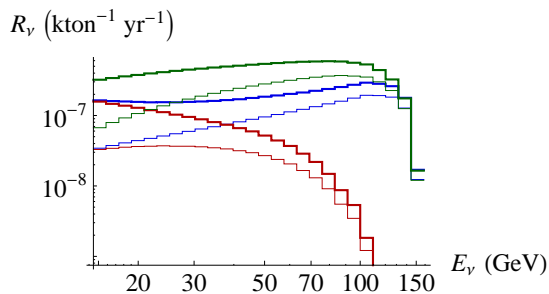}
  \caption{\label{fig:purebr_neu}\fulljust All-sky neutrino plus antineutrino detection rates for 150 GeV dark matter annihilation. The thick lines are for electron or muon flavor, and the thin lines show the tau flavor rate. At 70 Gev, the top two lines (green online) are for annihilation to $\tau^+\tau^-$ leptons, the middle two lines (blue online) show annihilation to $W^+W^-$ bosons, and the bottom two lines (red online) are for annihilation to $b\overline{b}$ quarks. \hfill\mbox{}}
\end{figure*}

Fig.~\ref{fig:purebr_neu} shows the neutrino event rates for annihilation into $W$~bosons, $b$~quarks, or $\tau$~leptons. The logarithmic GeV energy bin size used is
\[
  \Delta=\log_{10}\left(\frac{E_{i+1}}{E_i}\right)=0.04. 
\]
At these neutrino energies, the electron and muon neutrinos have indistinguishable nucleon scattering cross sections, which are larger than that for the tau neutrinos. Hence, the tau neutrino event rates are a little smaller.

Since $\tau$ leptons always decay to a primary neutrino, while $W$ bosons only decay directly to leptons some of the time, the $\nu$ production from $\tau$'s is more intense. The $b$ quarks do not produce primary neutrinos, and only have a lower energy neutrino spectrum from secondary chains. Thus, the flux of neutrinos from annihilations breaks the degeneracy between annihilation into $W^+W^-$ and $b\overline b$ that occured in the gamma-ray signal.

\begin{figure*}
  \includegraphics[width=0.4\textwidth]{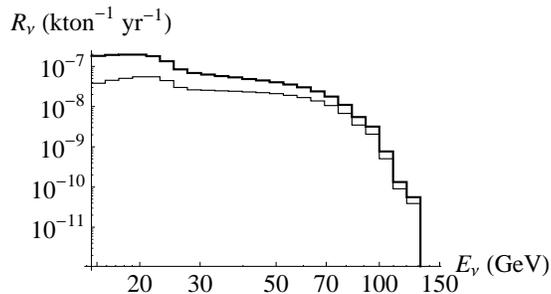}
  \caption{\label{fig:bml_neu}\fulljust All-sky neutrino plus antineutrino event rates for 150 GeV sneutrino dark matter that annihilates to two 135 GeV right-handed neutrinos (each flavor equally represented), each of which decays to a light neutrino and 120 GeV standard model higgs particle. \hfill\mbox{}}
\end{figure*}

Another class of models that results in interesting phenomenology for dark-matter-annihilation neutrinos is the $U(1)_{B-L}$ extension of the MSSM, described in Sec.~\ref{sec:B-L}, in the parameter spaces where the sneutrino $\tilde N$ is the dark matter particle. The neutrino detector rates for one example model are shown in Fig.~\ref{fig:bml_neu}. Here, the 150 GeV $\tilde N$ annihilates into 120 GeV right-handed neutrinos $N^c$ that decay to a standard model neutrino and higgs boson, the latter of which decays mostly to $b$'s and $\tau$'s. The $\tilde N$ annihilation in this model does have a slight p-wave component, and the s-wave cross section is $\sigma v=\unit{1.1\times10^{-26}}{\centi\meter\cubed\per\second}$, giving the correct thermal dark matter relic density. 

The secondary neutrinos produced from the higgs decay result in a broad, soft spectrum, whereas the neutrinos produced directly from $N^c$ decays produce a narrower peak at lower energies on the order of the mass difference between the $N^c$ and the higgs. Due to the higgs decays, there is also a gamma-ray component to the signal.

In the case where the $\tilde N$ dark matter is heavier (larger than twice the higgs mass), and the $N^c$ mass still slightly smaller than it, then the physical neutrino peak occurs closer to the dark matter mass energy. This will produce a hard spectrum with narrow peak from the primary neutrinos, and broad low-energy tail produced by the higgs decays.

\begin{figure*}
  \subfloat{\includegraphics[width=0.4\textwidth]
    {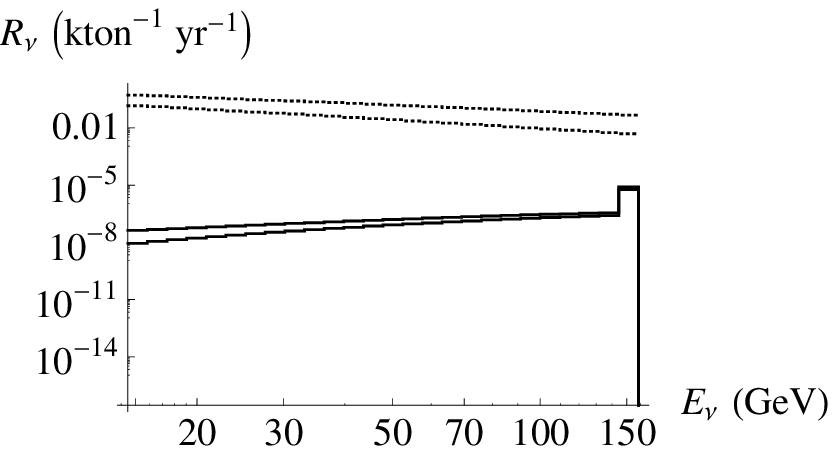}}\quad
  \subfloat{\includegraphics[width=0.4\textwidth]
    {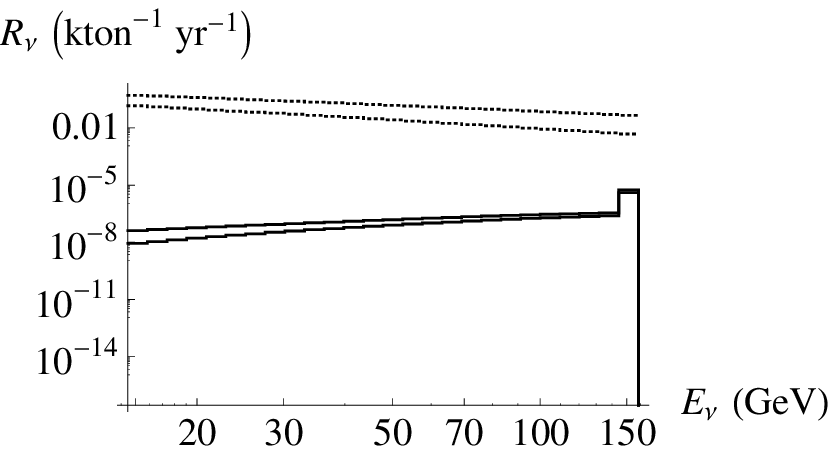}}\\
  \subfloat{\includegraphics[width=0.4\textwidth]
    {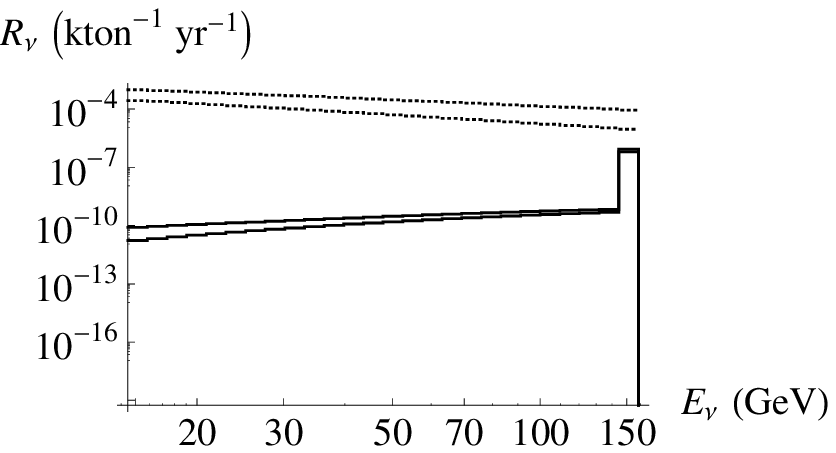}}\quad
  \subfloat{\includegraphics[width=0.4\textwidth]
    {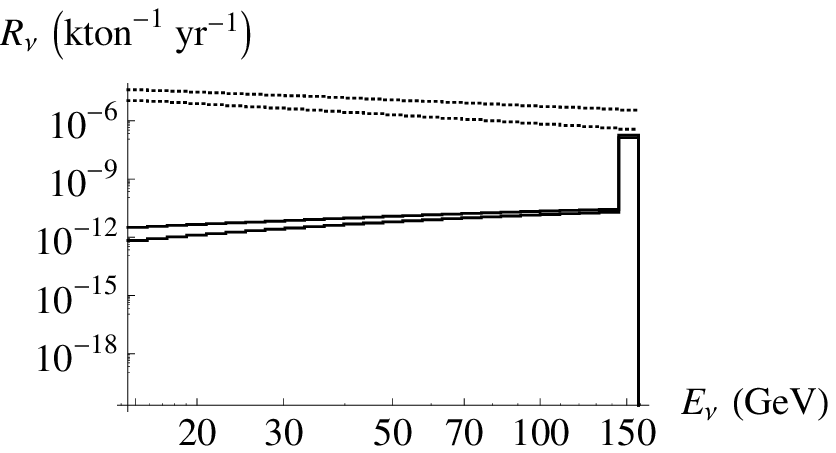}}
  \caption{\label{fig:line_neu}\fulljust Neutrino plus antineutrino event rates for 150 GeV dark matter annihilating to 2 prompt neutrinos $\nu$ with cross section $\sigma v=\unit{1.1\times10^{-26}}{\centi\meter\cubed\per\second}$ (solid lines), shown with the mean atmospheric neutrino plus antineutrino rates at the Kamioka site during low solar activity (dotted lines). For the atmospheric neutrinos, the upper line is the muon flavor, and the lower line is the electron flavor. For the annihilation neutrinos, the upper line shows the rate for electron flavor, as well as the rate for muon flavor. The lower line shows the rates for $\nu_\tau+\overline{\nu}_\tau$. Upper left: the mean neutrino rates from the whole sky. Upper right: rates when excluding the galactic core, $\psi>18^\circ$. Lower left: rates when focused on the galactic core, $\psi<5^\circ$. Lower right: rates when focused on the inner galactic core, $\psi<1^\circ$. \hfill\mbox{}}
\end{figure*}

Another intriguing scenario occurs when the dark matter annihilates solely to two light neutrinos $\nu$. In the context of the $B-L$ model previously described, this corresponds to the limit where the Higgs mass is small, negligible compared to the $\tilde N$ mass, and the mass difference between $\tilde N$ and $N^c$ is also very small. Then the spectrum of the produced light neutrinos is at the energy of the $\tilde N$, and the width of the spectrum is very small. This simple scenario results in a prominent neutrino line feature with no corresponding gamma-ray observations. At this energy scale of neutrino energies, the dominant astrophysical source is atmospheric neutrinos. The solid lines in Fig.~\ref{fig:line_neu} show the detector rates for annihilation of 150 GeV dark matter particles into prompt neutrinos, to each flavor equally, with a cross section of $\sigma v=\unit{1.1\times10^{-26}}{\centi\meter\cubed\per\second}$ with our modeled dark matter distribution. The upper line shows the electron flavor rates and muon flavor rates. The lower line is the tau flavor detection rate. Shown is the mean all-sky signal ($0^\circ<\psi<180^\circ$), an anticore signal ($\psi>18^\circ$), a core signal ($\psi<5^\circ$), and an inner core signal ($\psi<1^\circ$). The width of the spectral line feature is due to the velocity distribution of dark matter in the galactic halo, which is negligible compared to the energy resolution of viable detectors. Therefore, it is completely contained in the energy bin at the dark matter mass. The diffuse component is due to the redshifted extragalactic neutrinos. The dotted lines in the figure are the predicted mean atmospheric neutrino rates, as would be seen at the Kamioka site during minimum solar activity \cite{Honda:2006qj, *Honda:2011nf}. The upper line shows the $\nu_\mu+\overline{\nu}_\mu$ rates, and the lower line shows the $\nu_e+\overline{\nu}_e$ rates.

By comparing them with the previous neutrino rate plots, we see that the typical diffuse signals are well below the current measured atmospheric neutrino rates. Again, the situation likely improves with the consideration of halo substructure, and the background can also be reduced with respect to the signal by focusing on a nearby dark-matter-dense region of space, as we shall discuss for the prompt neutrino production example.

The prominence of the peak at the galactic core shows how a neutrino detector with high angular resolution may extract a spectral line feature by focusing on a dense region of space. Although the signal to background ratio improves with small solid angles of observation, the detection rates become forbiddingly small. With better energy resolution, an experiment can also gain a stronger signal in the spectral line scenario. Thinner energy bins have a higher spectral line height. The energy bin width at $E_\nu=m$ required for the bin height to be at the corresponding atmospheric neutrino rate when observing in solid angle $\Omega$ is approximately
\[
  \Delta E(m,\Omega)\approx\frac{1}{\Omega}\frac{\phi_G(m,\Omega)}{I_{\text{atm}}(m)-I_{\text{EG}}(m)}
\]
where $\phi_G(m,\Omega)$ is the flux of galactic annihilation neutrinos of energy at the dark matter particle mass $m$ originating within the solid angle $\Omega$, $I_{\text{atm}}(m)$ is the mean intensity of atmospheric neutrinos of energy $m$, and $I_{\text{EG}}(m)$ is the mean extragalactic annihilation neutrino intensity. The corresponding required logarithmic bin width is (assuming $\Delta E \ll m$)
\[
  \Delta\approx\frac{\Delta E}{m\ln10}.
\]

\begin{figure*}
  \subfloat{\includegraphics[width=0.33\textwidth]
    {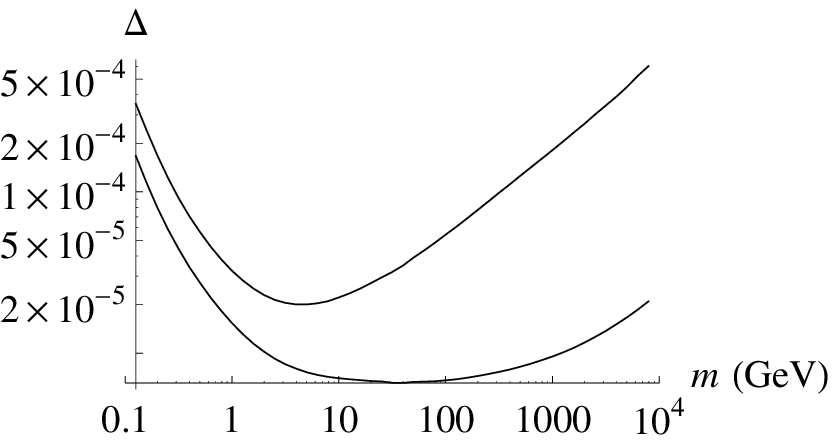}}
  \subfloat{\includegraphics[width=0.33\textwidth]
    {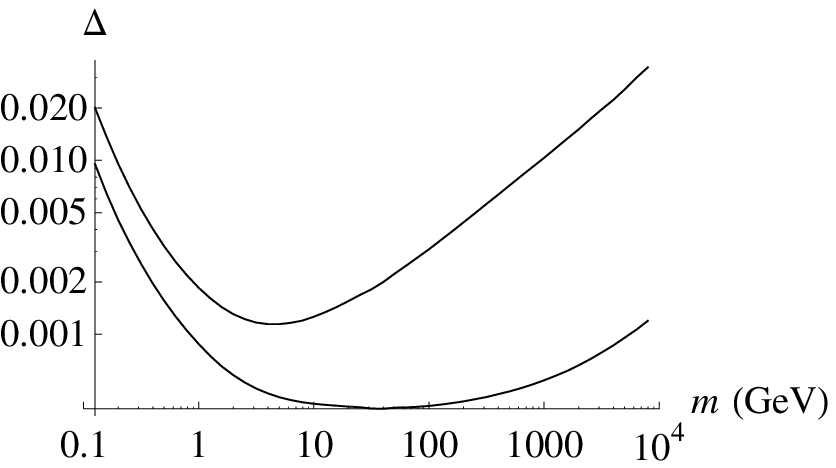}}
  \subfloat{\includegraphics[width=0.33\textwidth]
    {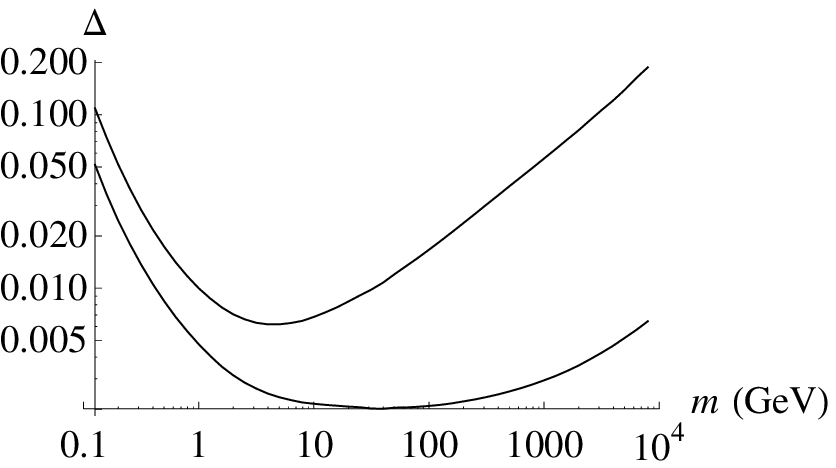}}
  \caption{\label{fig:neededbin}\fulljust The approximate logarithmic bin size required for the spectral line detector rate bin to reach the atmospheric neutrino detection rate. The upper line is for $\nu_e+\overline{\nu}_e$ and the lower line is for $\nu_\mu+\overline{\nu}_\mu$. Left: $\psi<180^\circ$. Center: $\psi<5^\circ$. Right: $\psi<1^\circ$. \hfill\mbox{}}
\end{figure*}

This approximate logarithmic energy bin size is shown in Fig.~\ref{fig:neededbin} for ranges of the dark matter mass, and for different solid angles centered on the galactic center. For comparison, Fig.~\ref{fig:line_neu} used $\Delta=0.04$. We see the energy scales where the spectral line is most hindered by the atmospheric neutrinos. At high dark matter mass, the electron neutrinos are much easier to see, since the electron atmospheric neutrinos are much less abundant than the muon atmospheric neutrinos.

The neutrino annihilation signal is also complementary to the detection of neutrinos from dark matter annihilations in the Sun. While the galactic and extragalactic signals depend on the dark matter's self-annihilation cross section, the solar annihilation signal is primarily dependent on the elastic nucleon-scattering cross section, provided that the annihilation rate and capture rate of dark matter particle by the Sun are in equilibrium. This fact means that it is still possible to probe models for which the elastic scattering cross section is too low for annihilations to be observed from the Sun. For example, in the context of $B-L$ models with $\tilde N$ dark matter, this occurs when there is a small mass splitting between the real and imaginary parts of $\tilde N$, and the dark matter becomes the lighter of the two \cite{Allahverdi:2009ae, *Allahverdi:2008jm, *Allahverdi:2009se}.
 
\section{Discussion}
\label{sec:discussion}

The possibility of indirect detection of dark matter annihilation through astrophysical observation of its products is an idea that improves our understanding of the particle nature of dark matter. It can already provide constraints on the s-wave part of its annihilation cross-section for a range of particle masses that annihilate according to some specified annihilation spectrum and assumptions about the dark matter distribution. These dark matter constraints are not available from any other kind of experiment.

Observation of annihilation products would provide valuable information about the dark matter self-interactions and distribution. The precise nature of the information that would be available to be extracted is dependent on the details of the dark matter properties, and on the nature of the data collected.

In this paper, we explored how the simultaneous observation of annihilation gamma-rays and neutrinos allows for more constrained conclusions of the dark matter properties and distribution. We gave sample calculations of the galactic and extragalctic contributions to signals from different particle physics models using a simple smooth halo model of dark matter distribution. It is interesting to consider gamma-rays and neutrinos because the direction of their source is constrained, and their signal contains contributions from both galactic and extragalactic sources. Observing these signals would improve the ability to reconstruct the particles produced in the annihilations, and the dark matter mass, by breaking degeneracies that exist in one signal. For example, dark matter of mass $\unit{150}{\giga\electronvolt}$ that annihilates directly into $W^+W^-$ bosons or $b\overline b$ quarks were seen to produce very similar gamma-ray spectra, but those different cases could be distinguishable in observations of cosmic neutrinos. For larger dark matter mass, the gamma-ray signals from annihilations into $W^+W^-$ will be more distinguishable from annihilations into $b\overline b$.

It is also possible for theories to be dominant in one kind of signal, and therefore undetectable in other channels. We discussed scenarios where annihilation could be neutrino-dominant. One intriguing case where this can happen is if dark matter is made up of supersymmetric partners of neutrinos, as we showed in the context of a low scale $B-L$ gauge theory, which can account for narrow spectral peaks in the cosmic neutrinos.

Once information about the dark matter particle mass is obtained, this theoretically constrains the Jeans mass scale responsible for the dark matter halo minimum mass. According to the halo model, the extragalactic signal is produced nearly democratically by all present scales of halo masses, when halo substructure is neglected. Since substructure has the most significant effect on large halos, one would predict the extragalactic signal to be dominated by the substructure of the most massive halos, which in turn is dependent on the scale of minimum halo mass.

From this analysis, we find that the magnitude of the annihilation products would provide information about the annihilation cross section, the concentration of subhalos within the most massive halos, and the density profile of their cores. The degeneracies that these quantities have on the intensity could be broken by information in a signal extracted verifiably from galactic sources, if enough independent observations can sufficiently constrain the dark matter distribution at those sources, perhaps via the angular distribution of the galactic signal to extract information about a universal core density profile. Neutrinos may be much more suited than gamma-rays for observing a signal from the dense galactic core, since there are fewer astrophysical sources of neutrinos at the relevant energies originating from that region.

It is evident that interpretation of observations will require precise, realistic theoretical predictions. It is possible to efficiently explore the effects of various aspects of large scale structure and particle physics with the development of semi-analytic descriptions of the structure. Through these methods, the important scales contributing to the predictions can be identified, and the robustness of the calculations against the uncertainties determined. It is thus that we may determine the constraints possible with modern and future experiments searching for indirect signals of dark matter annihilation.

\begin{acknowledgments}
The authors thank Carsten Rott for helpful comments. The work of R.A. is supported by the University of New Mexico Office of Research. The work of S.C. and B.D. is supported in part by DOE Grant DE-FG02-95ER40917.
\end{acknowledgments}

\appendix

\section{Algorithm for mean annihilation intensities across the core of NFW profiles}
\label{ap:nfwgalinten}

While the density cusp at the center of dark matter halos in the NFW model causes the observed dark matter annihilation intensity to be infinite in the direction toward the center of the halo, mean intensities over solid angles including the center are finite. Baryon cooling and, in larger halos, the presence of a supermassive black hole are some of the important effects that ultimately generate a more realistic core profile. To keep the dark matter distribution in this paper relatively simple, we do not attempt to model these effects, but assume the NFW profile throughout the halo.

In this appendix, we share our method for accurate calculation of annihilation intensity averages $\overline{I}(\psi_M)$ from observations over solid angles centered on the galactic center, with angular radius $\psi_M$. Refering to Eqs.~(\ref{eq:galinten})--(\ref{eq:galrmax}), we wish to evaluate
\[
  \overline{I}(E_\gamma,\psi_M)=\frac{\sigma v}{8\pi m^2}\frac{\der N_\gamma}{\der E_\gamma}(E_\gamma) \overline{J}(\psi_M)
\]
with
\[
  \overline{J}(\psi_M)=\frac{1}{1-\cos\psi_M}\int_0^{\psi_M}\der\psi\sin\psi\hat J(\psi).
\]
Let $x$ be the distance from the solar system, in units of the galactic halo scale radius $r_{s,G}$, along a line of sight at angle $\psi$ from the galactic center, and $x_\odot$ be distance of the solar system from the galactic center, also in units of $r_{s,G}$. Then
\[
  \frac{1-\cos\psi_M}{\rho_{s,G}^2r_{s,G}}\overline{J}(\psi_M)=\int_0^{\psi_M}\der\psi\sin\psi\int_0^{x_{\text{max}}(\psi)}\frac{\der x}{\left(x^2-2x_\odot x\cos\psi+x_\odot^2\right)\left(1+\sqrt{x^2-2x_\odot x\cos\psi+x_\odot^2}\right)^4}
\]
where
\[
  x_{\text{max}}(\psi)=x_\odot\cos\psi+\sqrt{c_G^2-(x_\odot\sin\psi)^2}
\]
expresses the halo boundary and, as before, the halo concentration is $c_G=R_{\text{vir},G}/r_{s,G}$. The integrand of $\overline{J}$ in these coordinates is irregular in the neighborhood of $\psi=0$ and $x=x_\odot$, precisely where the modeled density diverges at the halo center.

\begin{figure*}
  \includegraphics[width=0.4\textwidth]{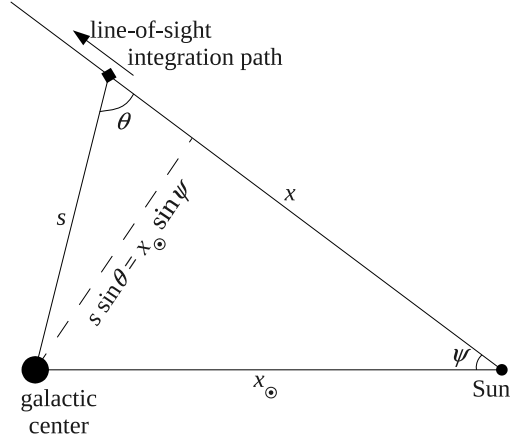}
  \caption{\label{fig:galcoords}\fulljust Galactic coordinates used for calculating the mean intensity due to dark matter annihilation in the smooth component of the galactic halo. \hfill\mbox{}}
\end{figure*}

The accurate evaluation of this expression is more easily attained when $x$ is replaced in favor of $\theta$, as pictured in Fig.~\ref{fig:galcoords}.
\[
  \sin\theta=\frac{x_\odot\sin\psi}{s}=\frac{x_\odot\sin\psi}{\sqrt{\left(x-x_\odot\cos\psi\right)^2+x_\odot^2\sin^2\psi}}
\]
In these coordinates,
\[
  \frac{1-\cos\psi_M}{\rho_{s,G}^2r_{s,G}}\overline{J}(\psi_M)=\frac{1}{x_\odot}\int_0^{\psi_M}\der\psi\int_{\theta_M(\psi)}^{\pi-\psi}\der\theta\left(\frac{\sin\theta}{\sin\theta+x_\odot\sin\psi}\right)^4,
\]
where
\[
  \theta_M(\psi)=\sin^{-1}\left(\frac{x_\odot}{c_G}\sin\psi\right).
\]
The inner $\theta$ integration is now well-defined and easy to numerically evaluate, except for when $\psi=0$, where the $\theta$ path of integration becomes degenerate, initially at the Sun having the value of $\pi$, and instantaneously becoming $0$ when crossing the galactic center. Since this degenerate point is an end of the $\psi$ integration, it is sufficient for numerical evaluation to consider the value of the inner integration in the limit as $\psi$ approaches $0$.

For $\psi\longrightarrow 0$, we have $\theta_M\longrightarrow x_\odot\psi/c_G\longrightarrow 0$, and the inner integral approaches
\[
  \int_{\theta_M(\psi)}^{\pi-\psi}\der\theta\left(\frac{\sin\theta}{\sin\theta+x_\odot\psi}\right)^4\longrightarrow\int_0^\pi\der\theta=\pi.
\]

For $\psi=\pi$, the $\theta$ integration path is simply of zero measure with $\theta=0$ constant along the path. Therefore, the inner integration vanishes for this value of $\psi$.

\bibliography{nuintensity}

\begin{thebibliography}{10}%
\makeatletter
\providecommand \@ifxundefined [1]{%
 \ifx #1\undefined \expandafter \@firstoftwo
 \else \expandafter \@secondoftwo
\fi
}%
\providecommand \@ifnum [1]{%
 \ifnum #1\expandafter \@firstoftwo
 \else \expandafter \@secondoftwo
\fi
}%
\providecommand \enquote [1]{``#1''}%
\providecommand \bibnamefont  [1]{#1}%
\providecommand \bibfnamefont [1]{#1}%
\providecommand \citenamefont [1]{#1}%
\providecommand\href[0]{\@sanitize\@href}%
\providecommand\@href[1]{\endgroup\@@startlink{#1}\endgroup\@@href}%
\providecommand\@@href[1]{#1\@@endlink}%
\providecommand \@sanitize [0]{\begingroup\catcode`\&12\catcode`\#12\relax}%
\@ifxundefined \pdfoutput {\@firstoftwo}{%
 \@ifnum{\z@=\pdfoutput}{\@firstoftwo}{\@secondoftwo}%
}{%
 \providecommand\@@startlink[1]{\leavevmode\special{html:<a href="#1">}}%
 \providecommand\@@endlink[0]{\special{html:</a>}}%
}{%
 \providecommand\@@startlink[1]{%
  \leavevmode
  \pdfstartlink
   attr{/Border[0 0 1 ]/H/I/C[0 1 1]}%
   user{/Subtype/Link/A<</Type/Action/S/URI/URI(#1)>>}%
  \relax
 }%
 \providecommand\@@endlink[0]{\pdfendlink}%
}%
\providecommand \url  [0]{\begingroup\@sanitize \@url }%
\providecommand \@url [1]{\endgroup\@href {#1}{\urlprefix}}%
\providecommand \urlprefix [0]{URL }%
\providecommand \Eprint[0]{\href }%
\@ifxundefined \urlstyle {%
  \providecommand \doi [1]{doi:\discretionary{}{}{}#1}%
}{%
  \providecommand \doi [0]{doi:\discretionary{}{}{}\begingroup
  \urlstyle{rm}\Url }%
}%
\providecommand \doibase [0]{http://dx.doi.org/}%
\providecommand \Doi[1]{\href{\doibase#1}}%
\providecommand \bibAnnote [3]{%
  \BibitemShut{#1}%
  \begin{quotation}\noindent
    \textsc{Key:}\ #2\\\textsc{Annotation:}\ #3%
  \end{quotation}%
}%
\providecommand \bibAnnoteFile [2]{%
  \IfFileExists{#2}{\bibAnnote {#1} {#2} {\input{#2}}}{}%
}%
\providecommand \typeout [0]{\immediate \write \m@ne }%
\providecommand \selectlanguage [0]{\@gobble}%
\providecommand \bibinfo [0]{\@secondoftwo}%
\providecommand \bibfield [0]{\@secondoftwo}%
\providecommand \translation [1]{[#1]}%
\providecommand \BibitemOpen[0]{}%
\providecommand \bibitemStop [0]{}%
\providecommand \bibitemNoStop [0]{.\EOS\space}%
\providecommand \EOS [0]{\spacefactor3000\relax}%
\providecommand \BibitemShut [1]{\csname bibitem#1\endcsname}%
\bibitem{Atwood:2009ez}%
  \BibitemOpen
  \bibfield{author}{%
  \bibinfo {author} {\bibfnamefont{W.~B.}\ \bibnamefont{Atwood}} \emph{et~al.}
  (\bibinfo {collaboration} {LAT}),\ }%
  \bibfield{journal}{%
  \Doi{10.1088/0004-637X/697/2/1071}{\bibinfo {journal} {Astrophys. J.}}\ }%
  \textbf{\bibinfo {volume} {697}},\ \bibinfo {pages} {1071} (\bibinfo {year}
  {2009}),\ \Eprint{http://arxiv.org/abs/0902.1089}{arXiv:0902.1089
  [astro-ph.IM]}%
  \bibAnnoteFile{NoStop}{Atwood:2009ez}%
\bibitem{Abbasi:2010ak}%
  \BibitemOpen
  \bibfield{author}{%
  \bibinfo {author} {\bibfnamefont{R.}~\bibnamefont{Abbasi}} \emph{et~al.}
  (\bibinfo {collaboration} {IceCube Collaboration}),\ }%
  \bibfield{journal}{%
  \Doi{10.1103/PhysRevD.82.072003}{\bibinfo {journal} {Phys.Rev.}}\ }%
  \textbf{\bibinfo {volume} {D82}},\ \bibinfo {pages} {072003} (\bibinfo {year}
  {2010}),\ \Eprint{http://arxiv.org/abs/1009.1442}{arXiv:1009.1442
  [astro-ph.CO]}%
  \bibAnnoteFile{NoStop}{Abbasi:2010ak}%
\bibitem{Baltz:2008wd}%
  \BibitemOpen
  \bibfield{author}{%
  \bibinfo {author} {\bibfnamefont{E.~A.}\ \bibnamefont{Baltz}} \emph{et~al.},\
  }%
  \bibfield{journal}{%
  \Doi{10.1088/1475-7516/2008/07/013}{\bibinfo {journal} {JCAP}}\ }%
  \textbf{\bibinfo {volume} {0807}},\ \bibinfo {pages} {013} (\bibinfo {year}
  {2008}),\ \Eprint{http://arxiv.org/abs/0806.2911}{arXiv:0806.2911
  [astro-ph]}%
  \bibAnnoteFile{NoStop}{Baltz:2008wd}%
\bibitem{Ackermann:2011bg}%
  \BibitemOpen
  \bibfield{author}{%
  \bibinfo {author} {\bibfnamefont{M.}~\bibnamefont{Ackermann}} \emph{et~al.},\
  }%
  \bibfield{journal}{%
  \Doi{10.1088/0004-637X/741/1/30}{\bibinfo {journal} {Astrophys. J.}}\ }%
  \textbf{\bibinfo {volume} {741}},\ \bibinfo {pages} {30} (\bibinfo {year}
  {2011}),\ \Eprint{http://arxiv.org/abs/1108.0501}{arXiv:1108.0501
  [astro-ph.CO]}%
  \bibAnnoteFile{NoStop}{Ackermann:2011bg}%
\bibitem{Abdo:2010ex}%
  \BibitemOpen
  \bibfield{author}{%
  \bibinfo {author} {\bibfnamefont{A.~A.}\ \bibnamefont{Abdo}} \emph{et~al.},\
  }%
  \bibfield{journal}{%
  \Doi{10.1088/0004-637X/712/1/147}{\bibinfo {journal} {Astrophys. J.}}\ }%
  \textbf{\bibinfo {volume} {712}},\ \bibinfo {pages} {147} (\bibinfo {year}
  {2010}),\ \Eprint{http://arxiv.org/abs/1001.4531}{arXiv:1001.4531
  [astro-ph.CO]}%
  \bibAnnoteFile{NoStop}{Abdo:2010ex}%
\bibitem{Abdo:2010dk}%
  \BibitemOpen
  \bibfield{author}{%
  \bibinfo {author} {\bibfnamefont{A.~A.}\ \bibnamefont{Abdo}} \emph{et~al.}
  (\bibinfo {collaboration} {Fermi-LAT}),\ }%
  \bibfield{journal}{%
  \Doi{10.1088/1475-7516/2010/04/014}{\bibinfo {journal} {JCAP}}\ }%
  \textbf{\bibinfo {volume} {1004}},\ \bibinfo {pages} {014} (\bibinfo {year}
  {2010}),\ \Eprint{http://arxiv.org/abs/1002.4415}{arXiv:1002.4415
  [astro-ph.CO]}%
  \bibAnnoteFile{NoStop}{Abdo:2010dk}%
\bibitem{Ackermann:2010rg}%
  \BibitemOpen
  \bibfield{author}{%
  \bibinfo {author} {\bibfnamefont{M.}~\bibnamefont{Ackermann}} \emph{et~al.},\
  }%
  \bibfield{journal}{%
  \Doi{10.1088/1475-7516/2010/05/025}{\bibinfo {journal} {JCAP}}\ }%
  \textbf{\bibinfo {volume} {1005}},\ \bibinfo {pages} {025} (\bibinfo {year}
  {2010}),\ \Eprint{http://arxiv.org/abs/1002.2239}{arXiv:1002.2239
  [astro-ph.CO]}%
  \bibAnnoteFile{NoStop}{Ackermann:2010rg}%
\bibitem{collaboration:2011wa}%
  \BibitemOpen
  \bibfield{author}{%
  \bibinfo {author} {\bibfnamefont{T.~F.-L.}\ \bibnamefont{collaboration}}}%
   (\bibinfo {year} {2011}),\
  \Eprint{http://arxiv.org/abs/1108.3546}{arXiv:1108.3546 [astro-ph.HE]}%
  \bibAnnoteFile{NoStop}{collaboration:2011wa}%
\bibitem{Cuoco:2010jb}%
  \BibitemOpen
  \bibfield{author}{%
  \bibinfo {author} {\bibfnamefont{A.}~\bibnamefont{Cuoco}}, \bibinfo {author}
  {\bibfnamefont{A.}~\bibnamefont{Sellerholm}}, \bibinfo {author}
  {\bibfnamefont{J.}~\bibnamefont{Conrad}},\ and\ \bibinfo {author}
  {\bibfnamefont{S.}~\bibnamefont{Hannestad}},\ }%
  \bibfield{journal}{%
  \Doi{10.1111/j.1365-2966.2011.18525.x}{\bibinfo {journal} {Mon. Not. Roy.
  Astron. Soc.}}\ }%
  \textbf{\bibinfo {volume} {414}},\ \bibinfo {pages} {2040} (\bibinfo {year}
  {2011}),\ \Eprint{http://arxiv.org/abs/1005.0843}{arXiv:1005.0843
  [astro-ph.HE]}%
  \bibAnnoteFile{NoStop}{Cuoco:2010jb}%
\bibitem{Ajello:2011dq}%
  \BibitemOpen
  \bibfield{author}{%
  \bibinfo {author} {\bibfnamefont{M.}~\bibnamefont{Ajello}} \emph{et~al.},\ }%
  \bibfield{journal}{%
  \Doi{10.1103/PhysRevD.84.032007}{\bibinfo {journal} {Phys. Rev.}}\ }%
  \textbf{\bibinfo {volume} {D84}},\ \bibinfo {pages} {032007} (\bibinfo {year}
  {2011}),\ \Eprint{http://arxiv.org/abs/1107.4272}{arXiv:1107.4272
  [astro-ph.HE]}%
  \bibAnnoteFile{NoStop}{Ajello:2011dq}%
\bibitem{Abbasi:2009vg}%
  \BibitemOpen
  \bibfield{author}{%
  \bibinfo {author} {\bibfnamefont{R.}~\bibnamefont{Abbasi}} \emph{et~al.}
  (\bibinfo {collaboration} {The IceCube collaboration}),\ }%
  \bibfield{journal}{%
  \Doi{10.1103/PhysRevD.81.057101}{\bibinfo {journal} {Phys.Rev.}}\ }%
  \textbf{\bibinfo {volume} {D81}},\ \bibinfo {pages} {057101} (\bibinfo {year}
  {2010}),\ \Eprint{http://arxiv.org/abs/0910.4480}{arXiv:0910.4480
  [astro-ph.CO]}%
  \bibAnnoteFile{NoStop}{Abbasi:2009vg}%
\bibitem{Abbasi:2011eq}%
  \BibitemOpen
  \bibfield{author}{%
  \bibinfo {author} {\bibfnamefont{R.}~\bibnamefont{Abbasi}} \emph{et~al.}
  (\bibinfo {collaboration} {IceCube Collaboration}),\ }%
  \bibfield{journal}{%
  \Doi{10.1103/PhysRevD.84.022004}{\bibinfo {journal} {Phys.Rev.}}\ }%
  \textbf{\bibinfo {volume} {D84}},\ \bibinfo {pages} {022004} (\bibinfo {year}
  {2011}),\ \Eprint{http://arxiv.org/abs/1101.3349}{arXiv:1101.3349
  [astro-ph.HE]}%
  \bibAnnoteFile{NoStop}{Abbasi:2011eq}%
\bibitem{Zavala:2009zr}%
  \BibitemOpen
  \bibfield{author}{%
  \bibinfo {author} {\bibfnamefont{J.}~\bibnamefont{Zavala}}, \bibinfo {author}
  {\bibfnamefont{V.}~\bibnamefont{Springel}},\ and\ \bibinfo {author}
  {\bibfnamefont{M.}~\bibnamefont{Boylan-Kolchin}},\ }%
  \bibfield{journal}{%
  \bibinfo {journal} {Mon. Not. Roy. Astron. Soc.}\ }%
  \textbf{\bibinfo {volume} {405}},\ \bibinfo {pages} {593} (\bibinfo {year}
  {2010}),\ \Eprint{http://arxiv.org/abs/0908.2428}{arXiv:0908.2428
  [astro-ph.CO]}%
  \bibAnnoteFile{NoStop}{Zavala:2009zr}%
\bibitem{Chamseddine:1982jx}%
  \BibitemOpen
  \bibfield{author}{%
  \bibinfo {author} {\bibfnamefont{A.~H.}\ \bibnamefont{Chamseddine}}, \bibinfo
  {author} {\bibfnamefont{R.~L.}\ \bibnamefont{Arnowitt}},\ and\ \bibinfo
  {author} {\bibfnamefont{P.}~\bibnamefont{Nath}},\ }%
  \bibfield{journal}{%
  \Doi{10.1103/PhysRevLett.49.970}{\bibinfo {journal} {Phys.Rev.Lett.}}\ }%
  \textbf{\bibinfo {volume} {49}},\ \bibinfo {pages} {970} (\bibinfo {year}
  {1982})%
  \bibAnnoteFile{NoStop}{Chamseddine:1982jx}%
\bibitem{Barbieri:1982eh}%
  \BibitemOpen
  \bibfield{author}{%
  \bibinfo {author} {\bibfnamefont{R.}~\bibnamefont{Barbieri}}, \bibinfo
  {author} {\bibfnamefont{S.}~\bibnamefont{Ferrara}},\ and\ \bibinfo {author}
  {\bibfnamefont{C.~A.}\ \bibnamefont{Savoy}},\ }%
  \bibfield{journal}{%
  \Doi{10.1016/0370-2693(82)90685-2}{\bibinfo {journal} {Phys.Lett.}}\ }%
  \textbf{\bibinfo {volume} {B119}},\ \bibinfo {pages} {343} (\bibinfo {year}
  {1982})%
  \bibAnnoteFile{NoStop}{Barbieri:1982eh}%
\bibitem{Hall:1983iz}%
  \BibitemOpen
  \bibfield{author}{%
  \bibinfo {author} {\bibfnamefont{L.~J.}\ \bibnamefont{Hall}}, \bibinfo
  {author} {\bibfnamefont{J.~D.}\ \bibnamefont{Lykken}},\ and\ \bibinfo
  {author} {\bibfnamefont{S.}~\bibnamefont{Weinberg}},\ }%
  \bibfield{journal}{%
  \Doi{10.1103/PhysRevD.27.2359}{\bibinfo {journal} {Phys.Rev.}}\ }%
  \textbf{\bibinfo {volume} {D27}},\ \bibinfo {pages} {2359} (\bibinfo {year}
  {1983})%
  \bibAnnoteFile{NoStop}{Hall:1983iz}%
\bibitem{Nath:1983aw}%
  \BibitemOpen
  \bibfield{author}{%
  \bibinfo {author} {\bibfnamefont{P.}~\bibnamefont{Nath}}, \bibinfo {author}
  {\bibfnamefont{R.~L.}\ \bibnamefont{Arnowitt}},\ and\ \bibinfo {author}
  {\bibfnamefont{A.~H.}\ \bibnamefont{Chamseddine}},\ }%
  \bibfield{journal}{%
  \Doi{10.1016/0550-3213(83)90145-1}{\bibinfo {journal} {Nucl.Phys.}}\ }%
  \textbf{\bibinfo {volume} {B227}},\ \bibinfo {pages} {121} (\bibinfo {year}
  {1983})%
  \bibAnnoteFile{NoStop}{Nath:1983aw}%
\bibitem{Nilles:1983ge}%
  \BibitemOpen
  \bibfield{author}{%
  \bibinfo {author} {\bibfnamefont{H.~P.}\ \bibnamefont{Nilles}},\ }%
  \bibfield{journal}{%
  \Doi{10.1016/0370-1573(84)90008-5}{\bibinfo {journal} {Phys.Rept.}}\ }%
  \textbf{\bibinfo {volume} {110}},\ \bibinfo {pages} {1} (\bibinfo {year}
  {1984})%
  \bibAnnoteFile{NoStop}{Nilles:1983ge}%
\bibitem{Griest:1990kh}%
  \BibitemOpen
  \bibfield{author}{%
  \bibinfo {author} {\bibfnamefont{K.}~\bibnamefont{Griest}}\ and\ \bibinfo
  {author} {\bibfnamefont{D.}~\bibnamefont{Seckel}},\ }%
  \bibfield{journal}{%
  \Doi{10.1103/PhysRevD.43.3191}{\bibinfo {journal} {Phys. Rev.}}\ }%
  \textbf{\bibinfo {volume} {D43}},\ \bibinfo {pages} {3191} (\bibinfo {year}
  {1991})%
  \bibAnnoteFile{NoStop}{Griest:1990kh}%
\bibitem{Chan:1997bi}%
  \BibitemOpen
  \bibfield{author}{%
  \bibinfo {author} {\bibfnamefont{K.~L.}\ \bibnamefont{Chan}}, \bibinfo
  {author} {\bibfnamefont{U.}~\bibnamefont{Chattopadhyay}},\ and\ \bibinfo
  {author} {\bibfnamefont{P.}~\bibnamefont{Nath}},\ }%
  \bibfield{journal}{%
  \Doi{10.1103/PhysRevD.58.096004}{\bibinfo {journal} {Phys. Rev.}}\ }%
  \textbf{\bibinfo {volume} {D58}},\ \bibinfo {pages} {096004} (\bibinfo {year}
  {1998}),\ \Eprint{http://arxiv.org/abs/hep-ph/9710473}{arXiv:hep-ph/9710473}%
  \bibAnnoteFile{NoStop}{Chan:1997bi}%
\bibitem{Feng:1999mn}%
  \BibitemOpen
  \bibfield{author}{%
  \bibinfo {author} {\bibfnamefont{J.~L.}\ \bibnamefont{Feng}}, \bibinfo
  {author} {\bibfnamefont{K.~T.}\ \bibnamefont{Matchev}},\ and\ \bibinfo
  {author} {\bibfnamefont{T.}~\bibnamefont{Moroi}},\ }%
  \bibfield{journal}{%
  \Doi{10.1103/PhysRevLett.84.2322}{\bibinfo {journal} {Phys. Rev. Lett.}}\ }%
  \textbf{\bibinfo {volume} {84}},\ \bibinfo {pages} {2322} (\bibinfo {year}
  {2000}),\ \Eprint{http://arxiv.org/abs/hep-ph/9908309}{arXiv:hep-ph/9908309}%
  \bibAnnoteFile{NoStop}{Feng:1999mn}%
\bibitem{Feng:1999zg}%
  \BibitemOpen
  \bibfield{author}{%
  \bibinfo {author} {\bibfnamefont{J.~L.}\ \bibnamefont{Feng}}, \bibinfo
  {author} {\bibfnamefont{K.~T.}\ \bibnamefont{Matchev}},\ and\ \bibinfo
  {author} {\bibfnamefont{T.}~\bibnamefont{Moroi}},\ }%
  \bibfield{journal}{%
  \Doi{10.1103/PhysRevD.61.075005}{\bibinfo {journal} {Phys. Rev.}}\ }%
  \textbf{\bibinfo {volume} {D61}},\ \bibinfo {pages} {075005} (\bibinfo {year}
  {2000}),\ \Eprint{http://arxiv.org/abs/hep-ph/9909334}{arXiv:hep-ph/9909334}%
  \bibAnnoteFile{NoStop}{Feng:1999zg}%
\bibitem{Baer:1995nq}%
  \BibitemOpen
  \bibfield{author}{%
  \bibinfo {author} {\bibfnamefont{H.}~\bibnamefont{Baer}}, \bibinfo {author}
  {\bibfnamefont{C.-h.}\ \bibnamefont{Chen}}, \bibinfo {author}
  {\bibfnamefont{F.}~\bibnamefont{Paige}},\ and\ \bibinfo {author}
  {\bibfnamefont{X.}~\bibnamefont{Tata}},\ }%
  \bibfield{journal}{%
  \Doi{10.1103/PhysRevD.52.2746}{\bibinfo {journal} {Phys. Rev.}}\ }%
  \textbf{\bibinfo {volume} {D52}},\ \bibinfo {pages} {2746} (\bibinfo {year}
  {1995}),\ \Eprint{http://arxiv.org/abs/hep-ph/9503271}{arXiv:hep-ph/9503271}%
  \bibAnnoteFile{NoStop}{Baer:1995nq}%
\bibitem{Baer:1995va}%
  \BibitemOpen
  \bibfield{author}{%
  \bibinfo {author} {\bibfnamefont{H.}~\bibnamefont{Baer}}, \bibinfo {author}
  {\bibfnamefont{C.-h.}\ \bibnamefont{Chen}}, \bibinfo {author}
  {\bibfnamefont{F.}~\bibnamefont{Paige}},\ and\ \bibinfo {author}
  {\bibfnamefont{X.}~\bibnamefont{Tata}},\ }%
  \bibfield{journal}{%
  \Doi{10.1103/PhysRevD.53.6241}{\bibinfo {journal} {Phys. Rev.}}\ }%
  \textbf{\bibinfo {volume} {D53}},\ \bibinfo {pages} {6241} (\bibinfo {year}
  {1996}),\ \Eprint{http://arxiv.org/abs/hep-ph/9512383}{arXiv:hep-ph/9512383}%
  \bibAnnoteFile{NoStop}{Baer:1995va}%
\bibitem{Baer:1998sz}%
  \BibitemOpen
  \bibfield{author}{%
  \bibinfo {author} {\bibfnamefont{H.}~\bibnamefont{Baer}}, \bibinfo {author}
  {\bibfnamefont{C.-h.}\ \bibnamefont{Chen}}, \bibinfo {author}
  {\bibfnamefont{M.}~\bibnamefont{Drees}}, \bibinfo {author}
  {\bibfnamefont{F.}~\bibnamefont{Paige}},\ and\ \bibinfo {author}
  {\bibfnamefont{X.}~\bibnamefont{Tata}},\ }%
  \bibfield{journal}{%
  \Doi{10.1103/PhysRevD.59.055014}{\bibinfo {journal} {Phys. Rev.}}\ }%
  \textbf{\bibinfo {volume} {D59}},\ \bibinfo {pages} {055014} (\bibinfo {year}
  {1999}),\ \Eprint{http://arxiv.org/abs/hep-ph/9809223}{arXiv:hep-ph/9809223}%
  \bibAnnoteFile{NoStop}{Baer:1998sz}%
\bibitem{Chatrchyan:2011qs}%
  \BibitemOpen
  \bibfield{author}{%
  \bibinfo {author} {\bibfnamefont{S.}~\bibnamefont{Chatrchyan}} \emph{et~al.}
  (\bibinfo {collaboration} {CMS}),\ }%
  \bibfield{journal}{%
  \Doi{10.1007/JHEP08(2011)156}{\bibinfo {journal} {JHEP}}\ }%
  \textbf{\bibinfo {volume} {08}},\ \bibinfo {pages} {156} (\bibinfo {year}
  {2011}),\ \Eprint{http://arxiv.org/abs/1107.1870}{arXiv:1107.1870 [hep-ex]}%
  \bibAnnoteFile{NoStop}{Chatrchyan:2011qs}%
\bibitem{Chatrchyan:2011zy}%
  \BibitemOpen
  \bibfield{author}{%
  \bibinfo {author} {\bibfnamefont{S.}~\bibnamefont{Chatrchyan}} \emph{et~al.}
  (\bibinfo {collaboration} {CMS})}%
   (\bibinfo {year} {2011}),\
  \Eprint{http://arxiv.org/abs/1109.2352}{arXiv:1109.2352 [hep-ex]}%
  \bibAnnoteFile{NoStop}{Chatrchyan:2011zy}%
\bibitem{Collaboration:2011iu}%
  \BibitemOpen
  \bibfield{author}{%
  \bibinfo {author} {\bibfnamefont{A.}~\bibnamefont{Collaboration}}}%
   (\bibinfo {year} {2011}),\
  \Eprint{http://arxiv.org/abs/1109.6606}{arXiv:1109.6606 [hep-ex]}%
  \bibAnnoteFile{NoStop}{Collaboration:2011iu}%
\bibitem{Aad:2011qa}%
  \BibitemOpen
  \bibfield{author}{%
  \bibinfo {author} {\bibfnamefont{G.}~\bibnamefont{Aad}} \emph{et~al.}
  (\bibinfo {collaboration} {Atlas})}%
   (\bibinfo {year} {2011}),\
  \Eprint{http://arxiv.org/abs/1110.2299}{arXiv:1110.2299 [hep-ex]}%
  \bibAnnoteFile{NoStop}{Aad:2011qa}%
\bibitem{Ellis:1998kh}%
  \BibitemOpen
  \bibfield{author}{%
  \bibinfo {author} {\bibfnamefont{J.~R.}\ \bibnamefont{Ellis}}, \bibinfo
  {author} {\bibfnamefont{T.}~\bibnamefont{Falk}},\ and\ \bibinfo {author}
  {\bibfnamefont{K.~A.}\ \bibnamefont{Olive}},\ }%
  \bibfield{journal}{%
  \Doi{10.1016/S0370-2693(98)01392-6}{\bibinfo {journal} {Phys. Lett.}}\ }%
  \textbf{\bibinfo {volume} {B444}},\ \bibinfo {pages} {367} (\bibinfo {year}
  {1998}),\ \Eprint{http://arxiv.org/abs/hep-ph/9810360}{arXiv:hep-ph/9810360}%
  \bibAnnoteFile{NoStop}{Ellis:1998kh}%
\bibitem{Ellis:1999mm}%
  \BibitemOpen
  \bibfield{author}{%
  \bibinfo {author} {\bibfnamefont{J.~R.}\ \bibnamefont{Ellis}}, \bibinfo
  {author} {\bibfnamefont{T.}~\bibnamefont{Falk}}, \bibinfo {author}
  {\bibfnamefont{K.~A.}\ \bibnamefont{Olive}},\ and\ \bibinfo {author}
  {\bibfnamefont{M.}~\bibnamefont{Srednicki}},\ }%
  \bibfield{journal}{%
  \Doi{10.1016/S0927-6505(99)00104-8}{\bibinfo {journal} {Astropart. Phys.}}\
  }%
  \textbf{\bibinfo {volume} {13}},\ \bibinfo {pages} {181} (\bibinfo {year}
  {2000}),\ \Eprint{http://arxiv.org/abs/hep-ph/9905481}{arXiv:hep-ph/9905481}%
  \bibAnnoteFile{NoStop}{Ellis:1999mm}%
\bibitem{Gomez:1999dk}%
  \BibitemOpen
  \bibfield{author}{%
  \bibinfo {author} {\bibfnamefont{M.~E.}\ \bibnamefont{Gomez}}, \bibinfo
  {author} {\bibfnamefont{G.}~\bibnamefont{Lazarides}},\ and\ \bibinfo {author}
  {\bibfnamefont{C.}~\bibnamefont{Pallis}},\ }%
  \bibfield{journal}{%
  \Doi{10.1103/PhysRevD.61.123512}{\bibinfo {journal} {Phys. Rev.}}\ }%
  \textbf{\bibinfo {volume} {D61}},\ \bibinfo {pages} {123512} (\bibinfo {year}
  {2000}),\ \Eprint{http://arxiv.org/abs/hep-ph/9907261}{arXiv:hep-ph/9907261}%
  \bibAnnoteFile{NoStop}{Gomez:1999dk}%
\bibitem{Gomez:2000sj}%
  \BibitemOpen
  \bibfield{author}{%
  \bibinfo {author} {\bibfnamefont{M.~E.}\ \bibnamefont{Gomez}}, \bibinfo
  {author} {\bibfnamefont{G.}~\bibnamefont{Lazarides}},\ and\ \bibinfo {author}
  {\bibfnamefont{C.}~\bibnamefont{Pallis}},\ }%
  \bibfield{journal}{%
  \Doi{10.1016/S0370-2693(00)00841-8}{\bibinfo {journal} {Phys. Lett.}}\ }%
  \textbf{\bibinfo {volume} {B487}},\ \bibinfo {pages} {313} (\bibinfo {year}
  {2000}),\ \Eprint{http://arxiv.org/abs/hep-ph/0004028}{arXiv:hep-ph/0004028}%
  \bibAnnoteFile{NoStop}{Gomez:2000sj}%
\bibitem{Lahanas:1999uy}%
  \BibitemOpen
  \bibfield{author}{%
  \bibinfo {author} {\bibfnamefont{A.~B.}\ \bibnamefont{Lahanas}}, \bibinfo
  {author} {\bibfnamefont{D.~V.}\ \bibnamefont{Nanopoulos}},\ and\ \bibinfo
  {author} {\bibfnamefont{V.~C.}\ \bibnamefont{Spanos}},\ }%
  \bibfield{journal}{%
  \Doi{10.1103/PhysRevD.62.023515}{\bibinfo {journal} {Phys. Rev.}}\ }%
  \textbf{\bibinfo {volume} {D62}},\ \bibinfo {pages} {023515} (\bibinfo {year}
  {2000}),\ \Eprint{http://arxiv.org/abs/hep-ph/9909497}{arXiv:hep-ph/9909497}%
  \bibAnnoteFile{NoStop}{Lahanas:1999uy}%
\bibitem{Arnowitt:2001yh}%
  \BibitemOpen
  \bibfield{author}{%
  \bibinfo {author} {\bibfnamefont{R.~L.}\ \bibnamefont{Arnowitt}}, \bibinfo
  {author} {\bibfnamefont{B.}~\bibnamefont{Dutta}},\ and\ \bibinfo {author}
  {\bibfnamefont{Y.}~\bibnamefont{Santoso}},\ }%
  \bibfield{journal}{%
  \Doi{10.1016/S0550-3213(01)00230-9}{\bibinfo {journal} {Nucl. Phys.}}\ }%
  \textbf{\bibinfo {volume} {B606}},\ \bibinfo {pages} {59} (\bibinfo {year}
  {2001}),\ \Eprint{http://arxiv.org/abs/hep-ph/0102181}{arXiv:hep-ph/0102181}%
  \bibAnnoteFile{NoStop}{Arnowitt:2001yh}%
\bibitem{Ellis:2001nx}%
  \BibitemOpen
  \bibfield{author}{%
  \bibinfo {author} {\bibfnamefont{J.~R.}\ \bibnamefont{Ellis}}, \bibinfo
  {author} {\bibfnamefont{K.~A.}\ \bibnamefont{Olive}},\ and\ \bibinfo {author}
  {\bibfnamefont{Y.}~\bibnamefont{Santoso}},\ }%
  \bibfield{journal}{%
  \Doi{10.1016/S0927-6505(02)00151-2}{\bibinfo {journal} {Astropart.Phys.}}\ }%
  \textbf{\bibinfo {volume} {18}},\ \bibinfo {pages} {395} (\bibinfo {year}
  {2003}),\ \Eprint{http://arxiv.org/abs/hep-ph/0112113}{arXiv:hep-ph/0112113
  [hep-ph]}%
  \bibAnnoteFile{NoStop}{Ellis:2001nx}%
\bibitem{Drees:1992am}%
  \BibitemOpen
  \bibfield{author}{%
  \bibinfo {author} {\bibfnamefont{M.}~\bibnamefont{Drees}}\ and\ \bibinfo
  {author} {\bibfnamefont{M.~M.}\ \bibnamefont{Nojiri}},\ }%
  \bibfield{journal}{%
  \Doi{10.1103/PhysRevD.47.376}{\bibinfo {journal} {Phys. Rev.}}\ }%
  \textbf{\bibinfo {volume} {D47}},\ \bibinfo {pages} {376} (\bibinfo {year}
  {1993}),\ \Eprint{http://arxiv.org/abs/hep-ph/9207234}{arXiv:hep-ph/9207234}%
  \bibAnnoteFile{NoStop}{Drees:1992am}%
\bibitem{Bell:2010ei}%
  \BibitemOpen
  \bibfield{author}{%
  \bibinfo {author} {\bibfnamefont{N.~F.}\ \bibnamefont{Bell}}, \bibinfo
  {author} {\bibfnamefont{J.~B.}\ \bibnamefont{Dent}}, \bibinfo {author}
  {\bibfnamefont{T.~D.}\ \bibnamefont{Jacques}},\ and\ \bibinfo {author}
  {\bibfnamefont{T.~J.}\ \bibnamefont{Weiler}},\ }%
  \bibfield{journal}{%
  \Doi{10.1103/PhysRevD.83.013001}{\bibinfo {journal} {Phys.Rev.}}\ }%
  \textbf{\bibinfo {volume} {D83}},\ \bibinfo {pages} {013001} (\bibinfo {year}
  {2011}),\ \Eprint{http://arxiv.org/abs/1009.2584}{arXiv:1009.2584 [hep-ph]}%
  \bibAnnoteFile{NoStop}{Bell:2010ei}%
\bibitem{Baer:1997ai}%
  \BibitemOpen
  \bibfield{author}{%
  \bibinfo {author} {\bibfnamefont{H.}~\bibnamefont{Baer}}\ and\ \bibinfo
  {author} {\bibfnamefont{M.}~\bibnamefont{Brhlik}},\ }%
  \bibfield{journal}{%
  \Doi{10.1103/PhysRevD.57.567}{\bibinfo {journal} {Phys. Rev.}}\ }%
  \textbf{\bibinfo {volume} {D57}},\ \bibinfo {pages} {567} (\bibinfo {year}
  {1998}),\ \Eprint{http://arxiv.org/abs/hep-ph/9706509}{arXiv:hep-ph/9706509}%
  \bibAnnoteFile{NoStop}{Baer:1997ai}%
\bibitem{Baer:2000jj}%
  \BibitemOpen
  \bibfield{author}{%
  \bibinfo {author} {\bibfnamefont{H.}~\bibnamefont{Baer}} \emph{et~al.},\ }%
  \bibfield{journal}{%
  \Doi{10.1103/PhysRevD.63.015007}{\bibinfo {journal} {Phys. Rev.}}\ }%
  \textbf{\bibinfo {volume} {D63}},\ \bibinfo {pages} {015007} (\bibinfo {year}
  {2000}),\ \Eprint{http://arxiv.org/abs/hep-ph/0005027}{arXiv:hep-ph/0005027}%
  \bibAnnoteFile{NoStop}{Baer:2000jj}%
\bibitem{Ellis:2001msa}%
  \BibitemOpen
  \bibfield{author}{%
  \bibinfo {author} {\bibfnamefont{J.~R.}\ \bibnamefont{Ellis}}, \bibinfo
  {author} {\bibfnamefont{T.}~\bibnamefont{Falk}}, \bibinfo {author}
  {\bibfnamefont{G.}~\bibnamefont{Ganis}}, \bibinfo {author}
  {\bibfnamefont{K.~A.}\ \bibnamefont{Olive}},\ and\ \bibinfo {author}
  {\bibfnamefont{M.}~\bibnamefont{Srednicki}},\ }%
  \bibfield{journal}{%
  \Doi{10.1016/S0370-2693(01)00541-X}{\bibinfo {journal} {Phys. Lett.}}\ }%
  \textbf{\bibinfo {volume} {B510}},\ \bibinfo {pages} {236} (\bibinfo {year}
  {2001}),\ \Eprint{http://arxiv.org/abs/hep-ph/0102098}{arXiv:hep-ph/0102098}%
  \bibAnnoteFile{NoStop}{Ellis:2001msa}%
\bibitem{Roszkowski:2001sb}%
  \BibitemOpen
  \bibfield{author}{%
  \bibinfo {author} {\bibfnamefont{L.}~\bibnamefont{Roszkowski}}, \bibinfo
  {author} {\bibfnamefont{R.}~\bibnamefont{Ruiz~de Austri}},\ and\ \bibinfo
  {author} {\bibfnamefont{T.}~\bibnamefont{Nihei}},\ }%
  \bibfield{journal}{%
  \bibinfo {journal} {JHEP}\ }%
  \textbf{\bibinfo {volume} {08}},\ \bibinfo {pages} {024} (\bibinfo {year}
  {2001}),\ \Eprint{http://arxiv.org/abs/hep-ph/0106334}{arXiv:hep-ph/0106334}%
  \bibAnnoteFile{NoStop}{Roszkowski:2001sb}%
\bibitem{Djouadi:2001yk}%
  \BibitemOpen
  \bibfield{author}{%
  \bibinfo {author} {\bibfnamefont{A.}~\bibnamefont{Djouadi}}, \bibinfo
  {author} {\bibfnamefont{M.}~\bibnamefont{Drees}},\ and\ \bibinfo {author}
  {\bibfnamefont{J.~L.}\ \bibnamefont{Kneur}},\ }%
  \bibfield{journal}{%
  \bibinfo {journal} {JHEP}\ }%
  \textbf{\bibinfo {volume} {08}},\ \bibinfo {pages} {055} (\bibinfo {year}
  {2001}),\ \Eprint{http://arxiv.org/abs/hep-ph/0107316}{arXiv:hep-ph/0107316}%
  \bibAnnoteFile{NoStop}{Djouadi:2001yk}%
\bibitem{Lahanas:2001yr}%
  \BibitemOpen
  \bibfield{author}{%
  \bibinfo {author} {\bibfnamefont{A.~B.}\ \bibnamefont{Lahanas}}\ and\
  \bibinfo {author} {\bibfnamefont{V.~C.}\ \bibnamefont{Spanos}},\ }%
  \bibfield{journal}{%
  \Doi{10.1007/s100520100861}{\bibinfo {journal} {Eur. Phys. J.}}\ }%
  \textbf{\bibinfo {volume} {C23}},\ \bibinfo {pages} {185} (\bibinfo {year}
  {2002}),\ \Eprint{http://arxiv.org/abs/hep-ph/0106345}{arXiv:hep-ph/0106345}%
  \bibAnnoteFile{NoStop}{Lahanas:2001yr}%
\bibitem{Mohapatra:1980qe}%
  \BibitemOpen
  \bibfield{author}{%
  \bibinfo {author} {\bibfnamefont{R.~N.}\ \bibnamefont{Mohapatra}}\ and\
  \bibinfo {author} {\bibfnamefont{R.}~\bibnamefont{Marshak}},\ }%
  \bibfield{journal}{%
  \Doi{10.1103/PhysRevLett.44.1316}{\bibinfo {journal} {Phys.Rev.Lett.}}\ }%
  \textbf{\bibinfo {volume} {44}},\ \bibinfo {pages} {1316} (\bibinfo {year}
  {1980})%
  \bibAnnoteFile{NoStop}{Mohapatra:1980qe}%
\bibitem{Mohapatra:1980qf}%
  \BibitemOpen
  \bibfield{author}{%
  \bibinfo {author} {\bibfnamefont{R.~N.}\ \bibnamefont{Mohapatra}}\ and\
  \bibinfo {author} {\bibfnamefont{R.}~\bibnamefont{Marshak}},\ }%
  \bibfield{journal}{%
  \Doi{10.1103/PhysRevLett.44.1644.2}{\bibinfo {journal} {Phys.Rev.Lett.}}\ }%
  \textbf{\bibinfo {volume} {44}},\ \bibinfo {pages} {1644} (\bibinfo {year}
  {1980})%
  \bibAnnoteFile{NoStop}{Mohapatra:1980qf}%
\bibitem{Minkowski:1977sc}%
  \BibitemOpen
  \bibfield{author}{%
  \bibinfo {author} {\bibfnamefont{P.}~\bibnamefont{Minkowski}},\ }%
  \bibfield{journal}{%
  \Doi{10.1016/0370-2693(77)90435-X}{\bibinfo {journal} {Phys.Lett.}}\ }%
  \textbf{\bibinfo {volume} {B67}},\ \bibinfo {pages} {421} (\bibinfo {year}
  {1977})%
  \bibAnnoteFile{NoStop}{Minkowski:1977sc}%
\bibitem{Yanagida:1979as}%
  \BibitemOpen
  \bibfield{author}{%
  \bibinfo {author} {\bibfnamefont{T.}~\bibnamefont{Yanagida}},\ }%
  in\ \emph{\bibinfo {booktitle} {{Workshop on Unified Theories}}}\ (\bibinfo
  {year} {1979})\ p.~\bibinfo {pages} {95},\ \Eprint{http://arxiv.org/abs/KEK
  Report 79-18}{KEK Report 79-18}%
  \bibAnnoteFile{NoStop}{Yanagida:1979as}%
\bibitem{Gell-Mann:1979}%
  \BibitemOpen
  \bibfield{author}{%
  \bibinfo {author} {\bibfnamefont{M.}~\bibnamefont{Gell-Mann}}, \bibinfo
  {author} {\bibfnamefont{P.}~\bibnamefont{Ramond}},\ and\ \bibinfo {author}
  {\bibfnamefont{R.}~\bibnamefont{Slansky}},\ }%
  in\ \emph{\bibinfo {booktitle} {{Supergravity}}},\ \bibinfo {editor} {edited
  by\ \bibinfo {editor} {\bibfnamefont{P.}~\bibnamefont{van Niewenhuizen}}\
  and\ \bibinfo {editor} {\bibfnamefont{D.}~\bibnamefont{Freedman}}}\ (\bibinfo
  {publisher} {North Holland},\ \bibinfo {address} {Amsterdam},\ \bibinfo
  {year} {1979})\ p.\ \bibinfo {pages} {315}%
  \bibAnnoteFile{NoStop}{Gell-Mann:1979}%
\bibitem{Glashow:1979}%
  \BibitemOpen
  \bibfield{author}{%
  \bibinfo {author} {\bibfnamefont{S.~L.}\ \bibnamefont{Glashow}},\ }%
  in\ \emph{\bibinfo {booktitle} {{1979 Cargese Summer Institute on Quarks and
  Leptons}}}\ (\bibinfo {publisher} {Plenum},\ \bibinfo {address} {New York},\
  \bibinfo {year} {1980})\ p.\ \bibinfo {pages} {687}%
  \bibAnnoteFile{NoStop}{Glashow:1979}%
\bibitem{Mohapatra:1979ia}%
  \BibitemOpen
  \bibfield{author}{%
  \bibinfo {author} {\bibfnamefont{R.~N.}\ \bibnamefont{Mohapatra}}\ and\
  \bibinfo {author} {\bibfnamefont{G.}~\bibnamefont{Senjanovic}},\ }%
  \bibfield{journal}{%
  \Doi{10.1103/PhysRevLett.44.912}{\bibinfo {journal} {Phys.Rev.Lett.}}\ }%
  \textbf{\bibinfo {volume} {44}},\ \bibinfo {pages} {912} (\bibinfo {year}
  {1980})%
  \bibAnnoteFile{NoStop}{Mohapatra:1979ia}%
\bibitem{Allahverdi:2009ae}%
  \BibitemOpen
  \bibfield{author}{%
  \bibinfo {author} {\bibfnamefont{R.}~\bibnamefont{Allahverdi}}, \bibinfo
  {author} {\bibfnamefont{B.}~\bibnamefont{Dutta}}, \bibinfo {author}
  {\bibfnamefont{K.}~\bibnamefont{Richardson-McDaniel}},\ and\ \bibinfo
  {author} {\bibfnamefont{Y.}~\bibnamefont{Santoso}},\ }%
  \bibfield{journal}{%
  \Doi{10.1016/j.physletb.2009.05.034}{\bibinfo {journal} {Phys.Lett.}}\ }%
  \textbf{\bibinfo {volume} {B677}},\ \bibinfo {pages} {172} (\bibinfo {year}
  {2009}),\ \Eprint{http://arxiv.org/abs/0902.3463}{arXiv:0902.3463 [hep-ph]}%
  \bibAnnoteFile{NoStop}{Allahverdi:2009ae}%
\bibitem{Allahverdi:2008jm}%
  \BibitemOpen
  \bibfield{author}{%
  \bibinfo {author} {\bibfnamefont{R.}~\bibnamefont{Allahverdi}}, \bibinfo
  {author} {\bibfnamefont{B.}~\bibnamefont{Dutta}}, \bibinfo {author}
  {\bibfnamefont{K.}~\bibnamefont{Richardson-McDaniel}},\ and\ \bibinfo
  {author} {\bibfnamefont{Y.}~\bibnamefont{Santoso}},\ }%
  \bibfield{journal}{%
  \Doi{10.1103/PhysRevD.79.075005}{\bibinfo {journal} {Phys.Rev.}}\ }%
  \textbf{\bibinfo {volume} {D79}},\ \bibinfo {pages} {075005} (\bibinfo {year}
  {2009}),\ \Eprint{http://arxiv.org/abs/0812.2196}{arXiv:0812.2196 [hep-ph]}%
  \bibAnnoteFile{NoStop}{Allahverdi:2008jm}%
\bibitem{Allahverdi:2009se}%
  \BibitemOpen
  \bibfield{author}{%
  \bibinfo {author} {\bibfnamefont{R.}~\bibnamefont{Allahverdi}}, \bibinfo
  {author} {\bibfnamefont{S.}~\bibnamefont{Bornhauser}}, \bibinfo {author}
  {\bibfnamefont{B.}~\bibnamefont{Dutta}},\ and\ \bibinfo {author}
  {\bibfnamefont{K.}~\bibnamefont{Richardson-McDaniel}},\ }%
  \bibfield{journal}{%
  \Doi{10.1103/PhysRevD.80.055026}{\bibinfo {journal} {Phys.Rev.}}\ }%
  \textbf{\bibinfo {volume} {D80}},\ \bibinfo {pages} {055026} (\bibinfo {year}
  {2009}),\ \Eprint{http://arxiv.org/abs/0907.1486}{arXiv:0907.1486 [hep-ph]}%
  \bibAnnoteFile{NoStop}{Allahverdi:2009se}%
\bibitem{Campbell:2010xc}%
  \BibitemOpen
  \bibfield{author}{%
  \bibinfo {author} {\bibfnamefont{S.}~\bibnamefont{Campbell}}, \bibinfo
  {author} {\bibfnamefont{B.}~\bibnamefont{Dutta}},\ and\ \bibinfo {author}
  {\bibfnamefont{E.}~\bibnamefont{Komatsu}},\ }%
  \bibfield{journal}{%
  \Doi{10.1103/PhysRevD.82.095007}{\bibinfo {journal} {Phys.Rev.}}\ }%
  \textbf{\bibinfo {volume} {D82}},\ \bibinfo {pages} {095007} (\bibinfo {year}
  {2010}),\ \Eprint{http://arxiv.org/abs/1009.3530}{arXiv:1009.3530 [hep-ph]}%
  \bibAnnoteFile{NoStop}{Campbell:2010xc}%
\bibitem{Campbell:2011kf}%
  \BibitemOpen
  \bibfield{author}{%
  \bibinfo {author} {\bibfnamefont{S.}~\bibnamefont{Campbell}}\ and\ \bibinfo
  {author} {\bibfnamefont{B.}~\bibnamefont{Dutta}},\ }%
  \bibfield{journal}{%
  \Doi{10.1103/PhysRevD.84.075004}{\bibinfo {journal} {Phys.Rev.}}\ }%
  \textbf{\bibinfo {volume} {D84}},\ \bibinfo {pages} {075004} (\bibinfo {year}
  {2011}),\ \Eprint{http://arxiv.org/abs/1106.4621}{arXiv:1106.4621
  [astro-ph.HE]}%
  \bibAnnoteFile{NoStop}{Campbell:2011kf}%
\bibitem{Sheth:1999su}%
  \BibitemOpen
  \bibfield{author}{%
  \bibinfo {author} {\bibfnamefont{R.~K.}\ \bibnamefont{Sheth}}, \bibinfo
  {author} {\bibfnamefont{H.~J.}\ \bibnamefont{Mo}},\ and\ \bibinfo {author}
  {\bibfnamefont{G.}~\bibnamefont{Tormen}},\ }%
  \bibfield{journal}{%
  \Doi{10.1046/j.1365-8711.2001.04006.x}{\bibinfo {journal} {Mon. Not. Roy.
  Astron. Soc.}}\ }%
  \textbf{\bibinfo {volume} {323}},\ \bibinfo {pages} {1} (\bibinfo {year}
  {2001}),\
  \Eprint{http://arxiv.org/abs/astro-ph/9907024}{arXiv:astro-ph/9907024}%
  \bibAnnoteFile{NoStop}{Sheth:1999su}%
\bibitem{Sheth:2001dp}%
  \BibitemOpen
  \bibfield{author}{%
  \bibinfo {author} {\bibfnamefont{R.~K.}\ \bibnamefont{Sheth}}\ and\ \bibinfo
  {author} {\bibfnamefont{G.}~\bibnamefont{Tormen}},\ }%
  \bibfield{journal}{%
  \Doi{10.1046/j.1365-8711.2002.04950.x}{\bibinfo {journal} {Mon. Not. Roy.
  Astron. Soc.}}\ }%
  \textbf{\bibinfo {volume} {329}},\ \bibinfo {pages} {61} (\bibinfo {year}
  {2002}),\
  \Eprint{http://arxiv.org/abs/astro-ph/0105113}{arXiv:astro-ph/0105113}%
  \bibAnnoteFile{NoStop}{Sheth:2001dp}%
\bibitem{Navarro:1996gj}%
  \BibitemOpen
  \bibfield{author}{%
  \bibinfo {author} {\bibfnamefont{J.~F.}\ \bibnamefont{Navarro}}, \bibinfo
  {author} {\bibfnamefont{C.~S.}\ \bibnamefont{Frenk}},\ and\ \bibinfo {author}
  {\bibfnamefont{S.~D.~M.}\ \bibnamefont{White}},\ }%
  \bibfield{journal}{%
  \Doi{10.1086/304888}{\bibinfo {journal} {Astrophys. J.}}\ }%
  \textbf{\bibinfo {volume} {490}},\ \bibinfo {pages} {493} (\bibinfo {year}
  {1997}),\
  \Eprint{http://arxiv.org/abs/astro-ph/9611107}{arXiv:astro-ph/9611107}%
  \bibAnnoteFile{NoStop}{Navarro:1996gj}%
\bibitem{Bullock:1999he}%
  \BibitemOpen
  \bibfield{author}{%
  \bibinfo {author} {\bibfnamefont{J.~S.}\ \bibnamefont{Bullock}}
  \emph{et~al.},\ }%
  \bibfield{journal}{%
  \Doi{10.1046/j.1365-8711.2001.04068.x}{\bibinfo {journal} {Mon. Not. Roy.
  Astron. Soc.}}\ }%
  \textbf{\bibinfo {volume} {321}},\ \bibinfo {pages} {559} (\bibinfo {year}
  {2001}),\
  \Eprint{http://arxiv.org/abs/astro-ph/9908159}{arXiv:astro-ph/9908159}%
  \bibAnnoteFile{NoStop}{Bullock:1999he}%
\bibitem{Komatsu:2010fb}%
  \BibitemOpen
  \bibfield{author}{%
  \bibinfo {author} {\bibfnamefont{E.}~\bibnamefont{Komatsu}} \emph{et~al.}
  (\bibinfo {collaboration} {WMAP Collaboration}),\ }%
  \bibfield{journal}{%
  \Doi{10.1088/0067-0049/192/2/18}{\bibinfo {journal} {Astrophys.J.Suppl.}}\ }%
  \textbf{\bibinfo {volume} {192}},\ \bibinfo {pages} {18} (\bibinfo {year}
  {2011}),\ \Eprint{http://arxiv.org/abs/1001.4538}{arXiv:1001.4538
  [astro-ph.CO]}%
  \bibAnnoteFile{NoStop}{Komatsu:2010fb}%
\bibitem{Eisenstein:1997jh}%
  \BibitemOpen
  \bibfield{author}{%
  \bibinfo {author} {\bibfnamefont{D.~J.}\ \bibnamefont{Eisenstein}}\ and\
  \bibinfo {author} {\bibfnamefont{W.}~\bibnamefont{Hu}},\ }%
  \bibfield{journal}{%
  \Doi{10.1086/306640}{\bibinfo {journal} {Astrophys.J.}}\ }%
  \textbf{\bibinfo {volume} {511}},\ \bibinfo {pages} {5} (\bibinfo {year}
  {1999}),\
  \Eprint{http://arxiv.org/abs/astro-ph/9710252}{arXiv:astro-ph/9710252
  [astro-ph]}%
  \bibAnnoteFile{NoStop}{Eisenstein:1997jh}%
\bibitem{Stecker:2005qs}%
  \BibitemOpen
  \bibfield{author}{%
  \bibinfo {author} {\bibfnamefont{F.~W.}\ \bibnamefont{Stecker}}, \bibinfo
  {author} {\bibfnamefont{M.~A.}\ \bibnamefont{Malkan}},\ and\ \bibinfo
  {author} {\bibfnamefont{S.~T.}\ \bibnamefont{Scully}},\ }%
  \bibfield{journal}{%
  \Doi{10.1086/506188}{\bibinfo {journal} {Astrophys. J.}}\ }%
  \textbf{\bibinfo {volume} {648}},\ \bibinfo {pages} {774} (\bibinfo {year}
  {2006}),\
  \Eprint{http://arxiv.org/abs/astro-ph/0510449}{arXiv:astro-ph/0510449}%
  \bibAnnoteFile{NoStop}{Stecker:2005qs}%
\bibitem{Stecker:2006eh}%
  \BibitemOpen
  \bibfield{author}{%
  \bibinfo {author} {\bibfnamefont{F.~W.}\ \bibnamefont{Stecker}}, \bibinfo
  {author} {\bibfnamefont{M.~A.}\ \bibnamefont{Malkan}},\ and\ \bibinfo
  {author} {\bibfnamefont{S.~T.}\ \bibnamefont{Scully}},\ }%
  \bibfield{journal}{%
  \Doi{10.1086/511738}{\bibinfo {journal} {Astrophys. J.}}\ }%
  \textbf{\bibinfo {volume} {658}},\ \bibinfo {pages} {1392} (\bibinfo {year}
  {2007}),\
  \Eprint{http://arxiv.org/abs/astro-ph/0612048}{arXiv:astro-ph/0612048}%
  \bibAnnoteFile{NoStop}{Stecker:2006eh}%
\bibitem{Barger:1987xg}%
  \BibitemOpen
  \bibfield{author}{%
  \bibinfo {author} {\bibfnamefont{V.~D.}\ \bibnamefont{Barger}}, \bibinfo
  {author} {\bibfnamefont{E.}~\bibnamefont{Glover}}, \bibinfo {author}
  {\bibfnamefont{K.}~\bibnamefont{Hikasa}}, \bibinfo {author}
  {\bibfnamefont{W.-Y.}\ \bibnamefont{Keung}}, \bibinfo {author}
  {\bibfnamefont{M.}~\bibnamefont{Olsson}}, \emph{et~al.},\ }%
  \bibfield{journal}{%
  \Doi{10.1103/PhysRevD.35.3366, 10.1103/PhysRevD.38.1632}{\bibinfo {journal}
  {Phys.Rev.}}\ }%
  \textbf{\bibinfo {volume} {D35}},\ \bibinfo {pages} {3366} (\bibinfo {year}
  {1987})%
  \bibAnnoteFile{NoStop}{Barger:1987xg}%
\bibitem{Feldman:2008xs}%
  \BibitemOpen
  \bibfield{author}{%
  \bibinfo {author} {\bibfnamefont{D.}~\bibnamefont{Feldman}}, \bibinfo
  {author} {\bibfnamefont{Z.}~\bibnamefont{Liu}},\ and\ \bibinfo {author}
  {\bibfnamefont{P.}~\bibnamefont{Nath}},\ }%
  \bibfield{journal}{%
  \Doi{10.1103/PhysRevD.79.063509}{\bibinfo {journal} {Phys. Rev.}}\ }%
  \textbf{\bibinfo {volume} {D79}},\ \bibinfo {pages} {063509} (\bibinfo {year}
  {2009}),\ \Eprint{http://arxiv.org/abs/0810.5762}{arXiv:0810.5762 [hep-ph]}%
  \bibAnnoteFile{NoStop}{Feldman:2008xs}%
\bibitem{Ibe:2008ye}%
  \BibitemOpen
  \bibfield{author}{%
  \bibinfo {author} {\bibfnamefont{M.}~\bibnamefont{Ibe}}, \bibinfo {author}
  {\bibfnamefont{H.}~\bibnamefont{Murayama}},\ and\ \bibinfo {author}
  {\bibfnamefont{T.~T.}\ \bibnamefont{Yanagida}},\ }%
  \bibfield{journal}{%
  \Doi{10.1103/PhysRevD.79.095009}{\bibinfo {journal} {Phys. Rev.}}\ }%
  \textbf{\bibinfo {volume} {D79}},\ \bibinfo {pages} {095009} (\bibinfo {year}
  {2009}),\ \Eprint{http://arxiv.org/abs/0812.0072}{arXiv:0812.0072 [hep-ph]}%
  \bibAnnoteFile{NoStop}{Ibe:2008ye}%
\bibitem{Backovic:2009rw}%
  \BibitemOpen
  \bibfield{author}{%
  \bibinfo {author} {\bibfnamefont{M.}~\bibnamefont{Backovic}}\ and\ \bibinfo
  {author} {\bibfnamefont{J.~P.}\ \bibnamefont{Ralston}},\ }%
  \bibfield{journal}{%
  \Doi{10.1103/PhysRevD.81.056002}{\bibinfo {journal} {Phys. Rev.}}\ }%
  \textbf{\bibinfo {volume} {D81}},\ \bibinfo {pages} {056002} (\bibinfo {year}
  {2010}),\ \Eprint{http://arxiv.org/abs/0910.1113}{arXiv:0910.1113 [hep-ph]}%
  \bibAnnoteFile{NoStop}{Backovic:2009rw}%
\bibitem{Hisano:2004ds}%
  \BibitemOpen
  \bibfield{author}{%
  \bibinfo {author} {\bibfnamefont{J.}~\bibnamefont{Hisano}}, \bibinfo {author}
  {\bibfnamefont{S.}~\bibnamefont{Matsumoto}}, \bibinfo {author}
  {\bibfnamefont{M.~M.}\ \bibnamefont{Nojiri}},\ and\ \bibinfo {author}
  {\bibfnamefont{O.}~\bibnamefont{Saito}},\ }%
  \bibfield{journal}{%
  \Doi{10.1103/PhysRevD.71.063528}{\bibinfo {journal} {Phys. Rev.}}\ }%
  \textbf{\bibinfo {volume} {D71}},\ \bibinfo {pages} {063528} (\bibinfo {year}
  {2005}),\ \Eprint{http://arxiv.org/abs/hep-ph/0412403}{arXiv:hep-ph/0412403}%
  \bibAnnoteFile{NoStop}{Hisano:2004ds}%
\bibitem{MarchRussell:2008yu}%
  \BibitemOpen
  \bibfield{author}{%
  \bibinfo {author} {\bibfnamefont{J.}~\bibnamefont{March-Russell}}, \bibinfo
  {author} {\bibfnamefont{S.~M.}\ \bibnamefont{West}}, \bibinfo {author}
  {\bibfnamefont{D.}~\bibnamefont{Cumberbatch}},\ and\ \bibinfo {author}
  {\bibfnamefont{D.}~\bibnamefont{Hooper}},\ }%
  \bibfield{journal}{%
  \Doi{10.1088/1126-6708/2008/07/058}{\bibinfo {journal} {JHEP}}\ }%
  \textbf{\bibinfo {volume} {0807}},\ \bibinfo {pages} {058} (\bibinfo {year}
  {2008}),\ \Eprint{http://arxiv.org/abs/0801.3440}{arXiv:0801.3440 [hep-ph]}%
  \bibAnnoteFile{NoStop}{MarchRussell:2008yu}%
\bibitem{ArkaniHamed:2008qn}%
  \BibitemOpen
  \bibfield{author}{%
  \bibinfo {author} {\bibfnamefont{N.}~\bibnamefont{Arkani-Hamed}}, \bibinfo
  {author} {\bibfnamefont{D.~P.}\ \bibnamefont{Finkbeiner}}, \bibinfo {author}
  {\bibfnamefont{T.~R.}\ \bibnamefont{Slatyer}},\ and\ \bibinfo {author}
  {\bibfnamefont{N.}~\bibnamefont{Weiner}},\ }%
  \bibfield{journal}{%
  \Doi{10.1103/PhysRevD.79.015014}{\bibinfo {journal} {Phys. Rev.}}\ }%
  \textbf{\bibinfo {volume} {D79}},\ \bibinfo {pages} {015014} (\bibinfo {year}
  {2009}),\ \Eprint{http://arxiv.org/abs/0810.0713}{arXiv:0810.0713 [hep-ph]}%
  \bibAnnoteFile{NoStop}{ArkaniHamed:2008qn}%
\bibitem{Lattanzi:2008qa}%
  \BibitemOpen
  \bibfield{author}{%
  \bibinfo {author} {\bibfnamefont{M.}~\bibnamefont{Lattanzi}}\ and\ \bibinfo
  {author} {\bibfnamefont{J.~I.}\ \bibnamefont{Silk}},\ }%
  \bibfield{journal}{%
  \Doi{10.1103/PhysRevD.79.083523}{\bibinfo {journal} {Phys. Rev.}}\ }%
  \textbf{\bibinfo {volume} {D79}},\ \bibinfo {pages} {083523} (\bibinfo {year}
  {2009}),\ \Eprint{http://arxiv.org/abs/0812.0360}{arXiv:0812.0360
  [astro-ph]}%
  \bibAnnoteFile{NoStop}{Lattanzi:2008qa}%
\bibitem{MarchRussell:2008tu}%
  \BibitemOpen
  \bibfield{author}{%
  \bibinfo {author} {\bibfnamefont{J.~D.}\ \bibnamefont{March-Russell}}\ and\
  \bibinfo {author} {\bibfnamefont{S.~M.}\ \bibnamefont{West}},\ }%
  \bibfield{journal}{%
  \Doi{10.1016/j.physletb.2009.04.010}{\bibinfo {journal} {Phys. Lett.}}\ }%
  \textbf{\bibinfo {volume} {B676}},\ \bibinfo {pages} {133} (\bibinfo {year}
  {2009}),\ \Eprint{http://arxiv.org/abs/0812.0559}{arXiv:0812.0559
  [astro-ph]}%
  \bibAnnoteFile{NoStop}{MarchRussell:2008tu}%
\bibitem{Iengo:2009ni}%
  \BibitemOpen
  \bibfield{author}{%
  \bibinfo {author} {\bibfnamefont{R.}~\bibnamefont{Iengo}},\ }%
  \bibfield{journal}{%
  \Doi{10.1088/1126-6708/2009/05/024}{\bibinfo {journal} {JHEP}}\ }%
  \textbf{\bibinfo {volume} {05}},\ \bibinfo {pages} {024} (\bibinfo {year}
  {2009}),\ \Eprint{http://arxiv.org/abs/0902.0688}{arXiv:0902.0688 [hep-ph]}%
  \bibAnnoteFile{NoStop}{Iengo:2009ni}%
\bibitem{Iengo:2009xf}%
  \BibitemOpen
  \bibfield{author}{%
  \bibinfo {author} {\bibfnamefont{R.}~\bibnamefont{Iengo}}}%
   (\bibinfo {year} {2009}),\
  \Eprint{http://arxiv.org/abs/0903.0317}{arXiv:0903.0317 [hep-ph]}%
  \bibAnnoteFile{NoStop}{Iengo:2009xf}%
\bibitem{Cassel:2009wt}%
  \BibitemOpen
  \bibfield{author}{%
  \bibinfo {author} {\bibfnamefont{S.}~\bibnamefont{Cassel}},\ }%
  \bibfield{journal}{%
  \Doi{10.1088/0954-3899/37/10/105009}{\bibinfo {journal} {J.Phys.G}}\ }%
  \textbf{\bibinfo {volume} {G37}},\ \bibinfo {pages} {105009} (\bibinfo {year}
  {2010}),\ \Eprint{http://arxiv.org/abs/0903.5307}{arXiv:0903.5307 [hep-ph]}%
  \bibAnnoteFile{NoStop}{Cassel:2009wt}%
\bibitem{Note1}%
  \BibitemOpen
  \bibinfo {note} {The $J$-factor is usually scaled to be in units of $R_\odot
  \rho _\odot $, where $\rho _\odot $ is the estimated local density. This is
  less convenient for comparison with the extragalactic signal, and therefore,
  we keep the $J$-factor with arbitrary units, and use the symbol $\protect
  \mathaccentV {hat}05E{J}$ to distinguish it from the more standard
  definition, to prevent confusion.}%
  \bibAnnoteFile{Stop}{Note1}%
\bibitem{Gondolo:2004sc}%
  \BibitemOpen
  \bibfield{author}{%
  \bibinfo {author} {\bibfnamefont{P.}~\bibnamefont{Gondolo}} \emph{et~al.},\
  }%
  \bibfield{journal}{%
  \Doi{10.1088/1475-7516/2004/07/008}{\bibinfo {journal} {JCAP}}\ }%
  \textbf{\bibinfo {volume} {0407}},\ \bibinfo {pages} {008} (\bibinfo {year}
  {2004}),\
  \Eprint{http://arxiv.org/abs/astro-ph/0406204}{arXiv:astro-ph/0406204}%
  \bibAnnoteFile{NoStop}{Gondolo:2004sc}%
\bibitem{Afshordi:2009hn}%
  \BibitemOpen
  \bibfield{author}{%
  \bibinfo {author} {\bibfnamefont{N.}~\bibnamefont{Afshordi}}, \bibinfo
  {author} {\bibfnamefont{R.}~\bibnamefont{Mohayaee}},\ and\ \bibinfo {author}
  {\bibfnamefont{E.}~\bibnamefont{Bertschinger}},\ }%
  \bibfield{journal}{%
  \Doi{10.1103/PhysRevD.81.101301}{\bibinfo {journal} {Phys.Rev.}}\ }%
  \textbf{\bibinfo {volume} {D81}},\ \bibinfo {pages} {101301} (\bibinfo {year}
  {2010}),\ \Eprint{http://arxiv.org/abs/0911.0414}{arXiv:0911.0414
  [astro-ph.CO]}%
  \bibAnnoteFile{NoStop}{Afshordi:2009hn}%
\bibitem{Klypin:2001xu}%
  \BibitemOpen
  \bibfield{author}{%
  \bibinfo {author} {\bibfnamefont{A.}~\bibnamefont{Klypin}}, \bibinfo {author}
  {\bibfnamefont{H.}~\bibnamefont{Zhao}},\ and\ \bibinfo {author}
  {\bibfnamefont{R.~S.}\ \bibnamefont{Somerville}},\ }%
  \bibfield{journal}{%
  \Doi{10.1086/340656}{\bibinfo {journal} {Astrophys.J.}}\ }%
  \textbf{\bibinfo {volume} {573}},\ \bibinfo {pages} {597} (\bibinfo {year}
  {2002}),\
  \Eprint{http://arxiv.org/abs/astro-ph/0110390}{arXiv:astro-ph/0110390
  [astro-ph]}%
  \bibAnnoteFile{NoStop}{Klypin:2001xu}%
\bibitem{Sjostrand:2007gs}%
  \BibitemOpen
  \bibfield{author}{%
  \bibinfo {author} {\bibfnamefont{T.}~\bibnamefont{Sjostrand}}, \bibinfo
  {author} {\bibfnamefont{S.}~\bibnamefont{Mrenna}},\ and\ \bibinfo {author}
  {\bibfnamefont{P.~Z.}\ \bibnamefont{Skands}},\ }%
  \bibfield{journal}{%
  \Doi{10.1016/j.cpc.2008.01.036}{\bibinfo {journal} {Comput. Phys. Commun.}}\
  }%
  \textbf{\bibinfo {volume} {178}},\ \bibinfo {pages} {852} (\bibinfo {year}
  {2008}),\ \Eprint{http://arxiv.org/abs/0710.3820}{arXiv:0710.3820 [hep-ph]}%
  \bibAnnoteFile{NoStop}{Sjostrand:2007gs}%
\bibitem{Gandhi:1998ri}%
  \BibitemOpen
  \bibfield{author}{%
  \bibinfo {author} {\bibfnamefont{R.}~\bibnamefont{Gandhi}}, \bibinfo {author}
  {\bibfnamefont{C.}~\bibnamefont{Quigg}}, \bibinfo {author}
  {\bibfnamefont{M.~H.}\ \bibnamefont{Reno}},\ and\ \bibinfo {author}
  {\bibfnamefont{I.}~\bibnamefont{Sarcevic}},\ }%
  \bibfield{journal}{%
  \Doi{10.1103/PhysRevD.58.093009}{\bibinfo {journal} {Phys.Rev.}}\ }%
  \textbf{\bibinfo {volume} {D58}},\ \bibinfo {pages} {093009} (\bibinfo {year}
  {1998}),\ \Eprint{http://arxiv.org/abs/hep-ph/9807264}{arXiv:hep-ph/9807264
  [hep-ph]}%
  \bibAnnoteFile{NoStop}{Gandhi:1998ri}%
\bibitem{Jeong:2010nt}%
  \BibitemOpen
  \bibfield{author}{%
  \bibinfo {author} {\bibfnamefont{Y.~S.}\ \bibnamefont{Jeong}}\ and\ \bibinfo
  {author} {\bibfnamefont{M.~H.}\ \bibnamefont{Reno}},\ }%
  \bibfield{journal}{%
  \Doi{10.1103/PhysRevD.82.033010}{\bibinfo {journal} {Phys.Rev.}}\ }%
  \textbf{\bibinfo {volume} {D82}},\ \bibinfo {pages} {033010} (\bibinfo {year}
  {2010}),\ \Eprint{http://arxiv.org/abs/1007.1966}{arXiv:1007.1966 [hep-ph]}%
  \bibAnnoteFile{NoStop}{Jeong:2010nt}%
\bibitem{Honda:2006qj}%
  \BibitemOpen
  \bibfield{author}{%
  \bibinfo {author} {\bibfnamefont{M.}~\bibnamefont{Honda}}, \bibinfo {author}
  {\bibfnamefont{T.}~\bibnamefont{Kajita}}, \bibinfo {author}
  {\bibfnamefont{K.}~\bibnamefont{Kasahara}}, \bibinfo {author}
  {\bibfnamefont{S.}~\bibnamefont{Midorikawa}},\ and\ \bibinfo {author}
  {\bibfnamefont{T.}~\bibnamefont{Sanuki}},\ }%
  \bibfield{journal}{%
  \Doi{10.1103/PhysRevD.75.043006}{\bibinfo {journal} {Phys.Rev.}}\ }%
  \textbf{\bibinfo {volume} {D75}},\ \bibinfo {pages} {043006} (\bibinfo {year}
  {2007}),\
  \Eprint{http://arxiv.org/abs/astro-ph/0611418}{arXiv:astro-ph/0611418
  [astro-ph]}%
  \bibAnnoteFile{NoStop}{Honda:2006qj}%
\bibitem{Honda:2011nf}%
  \BibitemOpen
  \bibfield{author}{%
  \bibinfo {author} {\bibfnamefont{M.}~\bibnamefont{Honda}}, \bibinfo {author}
  {\bibfnamefont{T.}~\bibnamefont{Kajita}}, \bibinfo {author}
  {\bibfnamefont{K.}~\bibnamefont{Kasahara}},\ and\ \bibinfo {author}
  {\bibfnamefont{S.}~\bibnamefont{Midorikawa}},\ }%
  \bibfield{journal}{%
  \Doi{10.1103/PhysRevD.83.123001}{\bibinfo {journal} {Phys.Rev.}}\ }%
  \textbf{\bibinfo {volume} {D83}},\ \bibinfo {pages} {123001} (\bibinfo {year}
  {2011}),\ \Eprint{http://arxiv.org/abs/1102.2688}{arXiv:1102.2688
  [astro-ph.HE]}%
  \bibAnnoteFile{NoStop}{Honda:2011nf}%
\end{thebibliography}%

\end{document}